\documentclass[pdftex,12pt,a4paper]{article}
\usepackage[affil-it]{authblk}
\usepackage[margin=0.7in]{geometry}
\usepackage{multicol}
\usepackage{enumerate}
\usepackage[pdftex]{graphicx}
\usepackage[font={normalsize,it}]{caption}
\usepackage[labelfont=bf]{caption}
\usepackage[margin=8pt]{subfig}
\usepackage{multirow}
\usepackage{amssymb,amsmath}
\usepackage{longtable}
\usepackage{float}
\usepackage{hyperref}
\usepackage[mathlines]{lineno}

\begin{document}
\title{Performance of Multiplexed XY Resistive Micromegas detectors in a high intensity beam}

\author[12]{D.~Banerjee}
\author[10]{V.~Burtsev}
\author[10]{A.~Chumakov}
\author[12]{D.~Cooke}
\author[12]{E.~Depero}
\author[5]{A.~V.~Dermenev}
\author[9]{S.~V.~Donskov}
\author[6]{F.~Dubinin}
\author[10]{R.~R.~Dusaev}
\author[12]{S.~Emmenegger}
\author[4]{A.~Fabich}
\author[2]{V.~N.~Frolov}
\author[8]{A.~Gardikiotis}
\author[5]{S.~N.~Gninenko}
\author[1]{M.~H\"osgen}
\author[5]{A.~E.~Karneyeu}
\author[1]{B.~Ketzer}
\author[5]{M.~M.~Kirsanov}
\author[3,6]{I.~V.~Konorov}
\author[7]{V.~A.~Kramarenko}
\author[11]{S.~V.~Kuleshov}
\author[10]{E. Levchenko}
\author[10,11]{V.~E.~Lyubovitskij}
\author[2]{V.~Lysan}
\author[10]{S. Mamon} 
\author[2]{V.~A.~Matveev}
\author[9]{Yu.~V.~Mikhailov}
\author[2]{V.~V.~Myalkovskiy}
\author[2]{V.~D.~Peshekhonov\footnote{Deceased}}
\author[2]{D.~V.~Peshekhonov}
\author[9]{V.~A.~Polyakov}
\author[12]{B.~Radics}
\author[12]{A.~Rubbia}
\author[9]{V.~D.~Samoylenko}
\author[6]{V.~O.~Tikhomirov}
\author[5]{D.~A.~Tlisov}
\author[5]{A.~N.~Toropin}
\author[10]{B.~Vasilishin}
\author[11]{G.~Vasquez Arenas}
\author[11]{P.~Ulloa}
\author[12]{P.~Crivelli\footnote{Corresponding author, crivelli@phys.ethz.ch}}


\affil[1]{\it Universit\"at Bonn, Helmholtz-Institut f\"ur Strahlen-und Kernphysik, 53115 Bonn, Germany} 
\affil[2]{\it  Joint Institute for Nuclear Research, 141980 Dubna, Russia}
\affil[3]{\it   Technische Universit\"at M\"unchen, Physik Department, 85748 Garching, Germany}
\affil[4]{\it CERN, European Organization for Nuclear Research, CH-1211 Geneva, Switzerland}
\affil[5]{\it Institute for Nuclear Research, 117312 Moscow, Russia}
\affil[6]{\it P.N. Lebedev Physics Institute, Moscow, Russia, 119 991 Moscow, Russia}
\affil[7]{\it Skobeltsyn Institute of Nuclear Physics, Lomonosov Moscow State University, Moscow, Russia}
\affil[8]{\it Physics Department, University of Patras, Patras, Greece} 
\affil[9]{\it State Scientific Center of the Russian Federation Institute for High Energy Physics of National Research Center 'Kurchatov Institute' (IHEP), 142281 Protvino, Russia}
\affil[10]{\it Tomsk Polytechnic University, 634050 Tomsk, Russia}
\affil[11]{\it Universidad T\'{e}cnica Federico Santa Mar\'{i}a, 2390123 Valpara\'{i}so, Chile}
\affil[12]{\it ETH Zurich, Institute for Particle Physics, CH-8093 Zurich, Switzerland}

\date{Dated: \today}
\maketitle
\clearpage
\begin{abstract}
We present the performance of multiplexed XY resistive Micromegas detectors tested in the CERN SPS 100 GeV/c electron beam at intensities up to 3.3 $\times$ 10$^5$ e$^- $/(s$\cdot$cm$^2$). 
So far, all studies with multiplexed Micromegas have only been reported for tests with radioactive sources and cosmic rays. 
The use of multiplexed modules in high intensity environments was not explored due to the effect of ambiguities in the reconstruction of the hit point caused by the multiplexing feature.
At the beam intensities analysed in this work and with a multiplexing 
factor of 5, more than 50\% level of ambiguity is introduced. Our results prove that by using the additional information of cluster size and integrated charge from the signal clusters induced on the XY strips, the ambiguities can be reduced to a level below 2\%.
The tested detectors are used in the CERN NA64 experiment for tracking the incoming particles bending in a magnetic field in order to reconstruct their momentum. 
The average hit detection efficiency of each module was found to be $\sim$ 96 $\%$ at the highest beam intensities. By using four modules a tracking resolution of 1.1 $\%$ was obtained with $\sim$ 85 $\%$ combined tracking efficiency.
\end{abstract}

\section{Introduction}

In the past years, a lot of effort has been invested in the development of microstrip gas detectors for particle tracking in various experiments (e.g \cite{cast}). Among those, Micromegas (MICRO-MEsh GASeous Structure)\cite{charpak} have found many applications in particle \cite{compass}, nuclear \cite{nuclear}  and astrophysics \cite{astro} for the detection of ionising particles. This relatively high-gain ($\sim$ 10$^{4}$)  gas detector  combines excellent spatial accuracy with a resolution below 100 $\mu$m \cite{Derre}, robustness, high rate capabilities, good timing resolution ($<$ 100 ns) and low material budget. Furthermore, this technology found applications in fire detectors \cite{fire} and muon tomography of volcanoes and pyramids \cite{appl_fut}.

Various improvements to this detector technology have continued since its first conception to make it more functional for applications in basic and applied research.  One of such  developments was the introduction of resistive strips to reduce the spark rate and, thus, limit the deadtime allowing Micromegas to operate in high flux environments \cite{rd51}. Several resistive Micromegas chambers with two-dimensional readout have been tested in the context of R\&D work for the ATLAS Muon System upgrade \cite{Wotschack}. 

The need of many experiments to have large scale tracking detectors with good spatial resolution is constantly growing. This implies a small strip pitch and hence a large number of readout channels. In order to make this more cost effective, an innovative technique for the readout, called genetic multiplexing, was developed  to sequentially connect multiple strips to individual readout channels thus reducing the required number of channels \cite{procureur}. 
However, this grouping may lead to fake combinations of ``ghost" clusters especially for high particle fluxes when pile-up is more likely and hence it was only tested for cosmic ray events \cite{mult_source}. 


In this paper we present the first measurements of resistive XY Micromegas modules multiplexed by a factor of 5 in a high particle flux. This test was done using the CERN SPS H4 high intensity secondary beamline in the context of the NA64 experiment \cite{NA64}. Our results show that fake combinations can be suppressed very efficiently by using the additional information of cluster size and integrated charge of signal clusters.
The experimental setup and the description of the Micromegas modules is presented in the following sections along-with its performance results.

\section{NA64}
\begin{figure}[H]
\centering
\includegraphics[width=1.\textwidth]{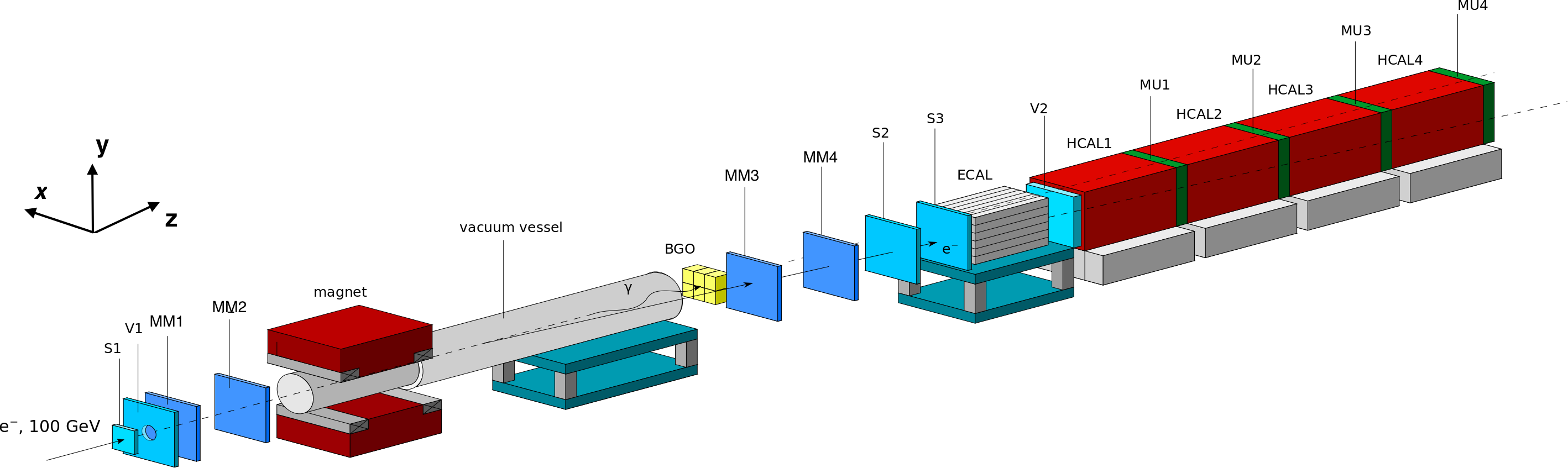}
\caption{NA64 detailed setup showing all sub-detectors (taken from \cite{PRL}).}
\label{setup_cartoon}
\end{figure}

The CERN NA64 experiment combines the active beam dump technique with missing energy measurement searching for invisible decays of massive $A'$ produced in the reaction: 
\begin{equation}
e^-Z \rightarrow e^-ZA'
\end{equation}
of electrons scattering off a nuclei of charge $Z$, with a mixing strength 10$^{-6} < \epsilon <$ 10$^{-3}$ and masses $M_{A'} $ in the sub-GeV range \cite{eps}. The secondary beam is produced by dumping the SPS 400 GeV protons at the Fixed Target T2 of the CERN North Area and transported to the detector in the evacuated H4 beamline tuned to a freely adjustable beam momentum from 10 up to 300 GeV/c. The detailed setup of the experiment is shown in Fig. \ref{setup_cartoon}.
The 100 GeV/c electron beam entering the setup is triggered by the coincidence of scintillators S1-S3. The typical maximal beam intensity  used for NA64 is of the order of 5 $\times$ 10$^6$ e$^-$ for a SPS spill with 10$^{12}$ protons on target \cite{beamline}. Each spill has a duration of 4.8 s and the beam has a diameter $\sim$ 2 cm (FWHM). The maximal beam intensity thus achieved is $\sim$  3.3 $\times$ 10$^5$ e$^-$/(s $\cdot$ cm$^2$). The beam is dumped on the electromagnetic calorimeter (ECAL), a sandwich of lead and scintillators (corresponding to 40 radiation lengths, $X_0$), to produce massive $A'$ through scattering with the heavy nuclei. In case of a $A'$ production, a fraction of the energy (the chosen threshold of the experiment is 0.5$\cdot$E$_0$ where E$_0$ is the beam energy) will be deposited in the ECAL and the rest will be carried away by the $A'$ without any interaction downstream of the ECAL. The signature for a signal will be missing energy in the ECAL and no activity in the VETO (V2, three plastic scintillator planes) and the four hadronic calorimeter modules (HCAL 1-4, a sandwich of iron and scintillators). The main sources of background for this search come from
\begin{enumerate}
\item electrons in the low energy tail of the beam mistaken as a high energy one depositing all its energy in the ECAL,
\item beam hadrons producing neutrals that carry away an energy larger than the ECAL threshold,
\item muons producing a low energy photon or delta electron with energy smaller than  0.5$\cdot$E$_0$ in the ECAL, which is absorbed in the calorimeter while the muon penetrates the rest of the detector without being detected.
\end{enumerate} 
A detailed description of all the expected background sources is presented in \cite{NA64}.  Beam hadrons and muons are suppressed at a level of 10$^{-5}$ by tagging the synchrotron radiation of the incoming particles deflected in the magnetic field \cite{sync}. To suppress low energy electrons, a spectrometer is required for NA64 to track the incoming particles and reconstruct their momentum in a magnetic field \cite{beam}. Any charged particle with momentum $p$ entering a magnetic field, $B$, is deflected by the field with the curvature radius of the trajectory $r = p/(qB)$, where $q$ is the charge of the particle. Four Multiplexed XY Resistive Micromegas detectors (MM1-MM4) were built for this reconstruction. The magnetic field used during the experiment is $\sim$ 7 T.m  in a 4.8 m long magnet. Two modules were positioned before the magnet $\sim$ 1 m apart and two downstream $\sim$ 12 m from the end of the magnet before the ECAL. MM3 and MM4 were placed $\sim$ 2 m from each other (see Fig. \ref{setup_cartoon}). 
\section{The Micromegas modules}

\subsection{Principle of Operation}
\begin{figure}[H]
\centering
\includegraphics[scale=0.5]{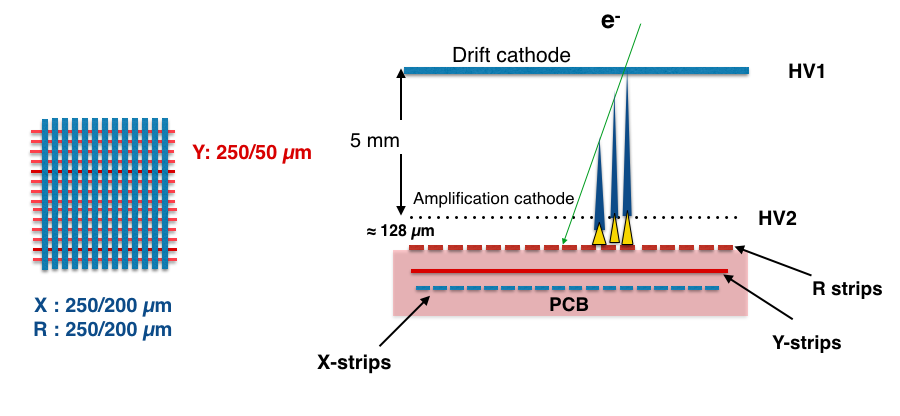}
\caption{Left: Sketch of the strip dimensions of the Micromegas modules. The pitch of the strip layers is 250 $\mu$m. Right: Principle of operation of a Micromegas Detector.}
\label{micro_principle}
\end{figure}

Micromegas detectors are two-stage parallel plate avalanche chambers with a narrow amplification gap and a wider drift gap as shown in the right picture of Fig. \ref{micro_principle}. Our modules have a drift gap of 5 mm separated from the amplification gap by a Nickel electro-formed micro-mesh. The drift cathode is made of a copper mesh and the amplification gap of 128 $\mu$m is defined by photo-resistive pillars 300 $\mu$m in diameter, equally spaced by 5 mm. The gas chambers are filled with mixtures of Ar and a quenching gas.
A charged particle entering the detector ionises the gas producing electrons that drift towards the micro-mesh under the electric field of the drift cathode, $\sim$ 0.6 kV/cm, wherein they enter the amplification region producing an avalanche of secondary electrons under the high amplification field $\sim$ 50 kV/cm. The signal induced on the X and Y strips is read by the front-end chips.  

The produced electrons in the amplification gap can also cause excitation of the gas molecules which return to the ground state via emission of UV photons \cite{51,52}. These UV photons can release new electrons from the gas molecules by photo-electric effect and eventually result in detector breakdown. Therefore, molecular gases with absorption bands in the UV range \cite{55} \cite{56} are mixed with the noble gas (7$\%$ CO$_2$ with 93$\%$ Ar in our case) to act as a ``quencher" to absorb these UV photons. 
\subsection{Design of NA64 Micromegas detectors}
\begin{figure}[H]
\centering
\includegraphics[scale=0.55]{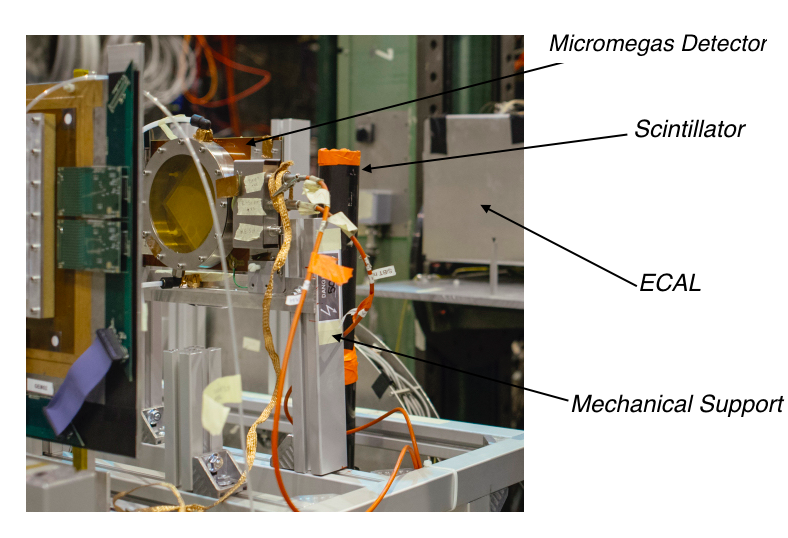}
\caption{Micromegas detector placed in the CERN SPS H4 Beam line.}
\label{MM_det}
\end{figure}

The results from the R\&D work performed for the ATLAS Muon System upgrade \cite{atlas_up} on several 2D Micromegas chambers with spark protection guided our design of the strip widths and pitch for the NA64 modules. Our resistive detectors were produced at the CERN  EP-DT-EF workshop. The readout strips are multiplexed by a factor 5 and the resistive strips (R strips) of resistance 50 M$\Omega$ are placed parallel to the X-strips as shown in the left picture of Fig.\ref{micro_principle}. The R and X strips have the same width of 200 $\mu$m with the Y strips placed after the R strips and perpendicular to the X-strips with a width of 50 $\mu$m. The pitch of all the strip layers is 250 $\mu$m. 
The active area is 8 cm $\times$ 8 cm, with 320 strips each for the X and Y coordinates. The readout is done with a 128 channel APV chip \cite{apv} as for the COMPASS GEM and Micromegas detectors \cite{Ketzer}. A multiplexing factor of 5 allows to have one chip per detector to read all 640 strips on both X and Y plane.
Fig. \ref{MM_det} shows one of the Micromegas modules setup at the CERN SPS H4 beam line. 

\subsection{Multiplexing}
The genetic multiplexing algorithm \cite{procureur} exploits the fact that a single particle entering the detector produces isolated tracks. The detected hits therefore occupy only a few neighbouring readout strips in a small area. This allows to group several strips to single electronic channels, thus reducing the size and cost of electronics. In this mapping construction there is only one set of two consecutive strips corresponding to a given set of two electronic channels. Using this scheme the theoretical number of readout strips that can be read by $p$ electronic channels is given by the maximum number of unordered doublets as
\begin{align*}
n_{max} &= \frac{p \times (p-1)}{2}+1 \quad \text{p=odd}\\
n_{max} &= \frac{p \times (p-2)}{2}+2 \quad \text{p=even}
\end{align*}
\paragraph{}
Therefore in principle the maximum number of strips that can be multiplexed to be read by the 128 channel APV chip is $\sim$ 8000. For the NA64 modules this multiplexing factor was reduced to 5 (corresponding to 640 strips per module) in order to limit ambiguities expected at high intensities. The multiplexing formula used to obtain the channel-strip ($c(s)$) mapping per plane where $c(s)$ is the channel corresponding to strip $s$ is:
\begin{equation}
c(s) = {\rm mod}(s\times({\rm floor}(s/p)\times m+1), p)
\end{equation}
where $p$ is the number of electronic channels = 64, $m$=6 and mod and floor are the modulo and the rounding down functions. m gives the maximum cluster size which does not lead to repetition of at least two consecutive strip connections. The above equation is, however, only valid when $p$ and $m+1$ does not share a common prime factor.
\paragraph{}
\subsection{Signal Cluster Reconstruction}
\label{sample}
\begin{figure}[t]
\centering
\includegraphics[scale=0.55]{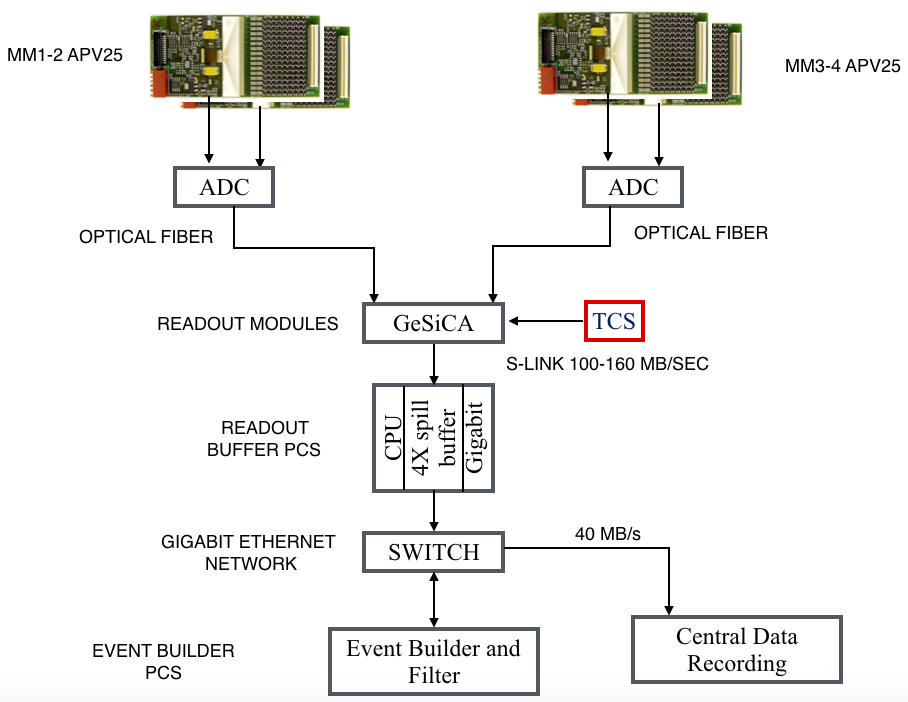}
\caption{General schematic of the NA64 DAQ.}
\label{DAQ_scheme}
\end{figure}

When the trigger from the experiment is received, the APV25 chips output for each channel three analog charge samples separated by 25 ns. Those signals are digitized by the ADC and are read by the common DAQ \cite{DAQ} of the experiment as shown in the general schematic of Fig. \ref{DAQ_scheme}. 
The pedestal distribution of the electronic channels, obtained from the weighted sum of the three charge samples/channel, are recorded in absence of the beam. A hit on a strip is defined when the weighted sum of the three samples are at least 3 standard deviations above the mean pedestal level. The maximum of the three samples is then recorded as the charge information of the hit strip. 
 When a charged particle enters the detector it is expected to leave a signal on consecutive strips due to its drift in the gas and the consequent charge spread. A signal cluster is defined when at least two neighbouring strips are hit. To reconstruct the position of a signal cluster, the electronic channels are  mapped to the multiplexed strips. The hit position on each plane (X and Y) is calculated from the weighted average of the detected charge on the strips.

\begin{figure}[H]
\begin{minipage}{.5\textwidth}
  \centering
\includegraphics[scale=0.47]{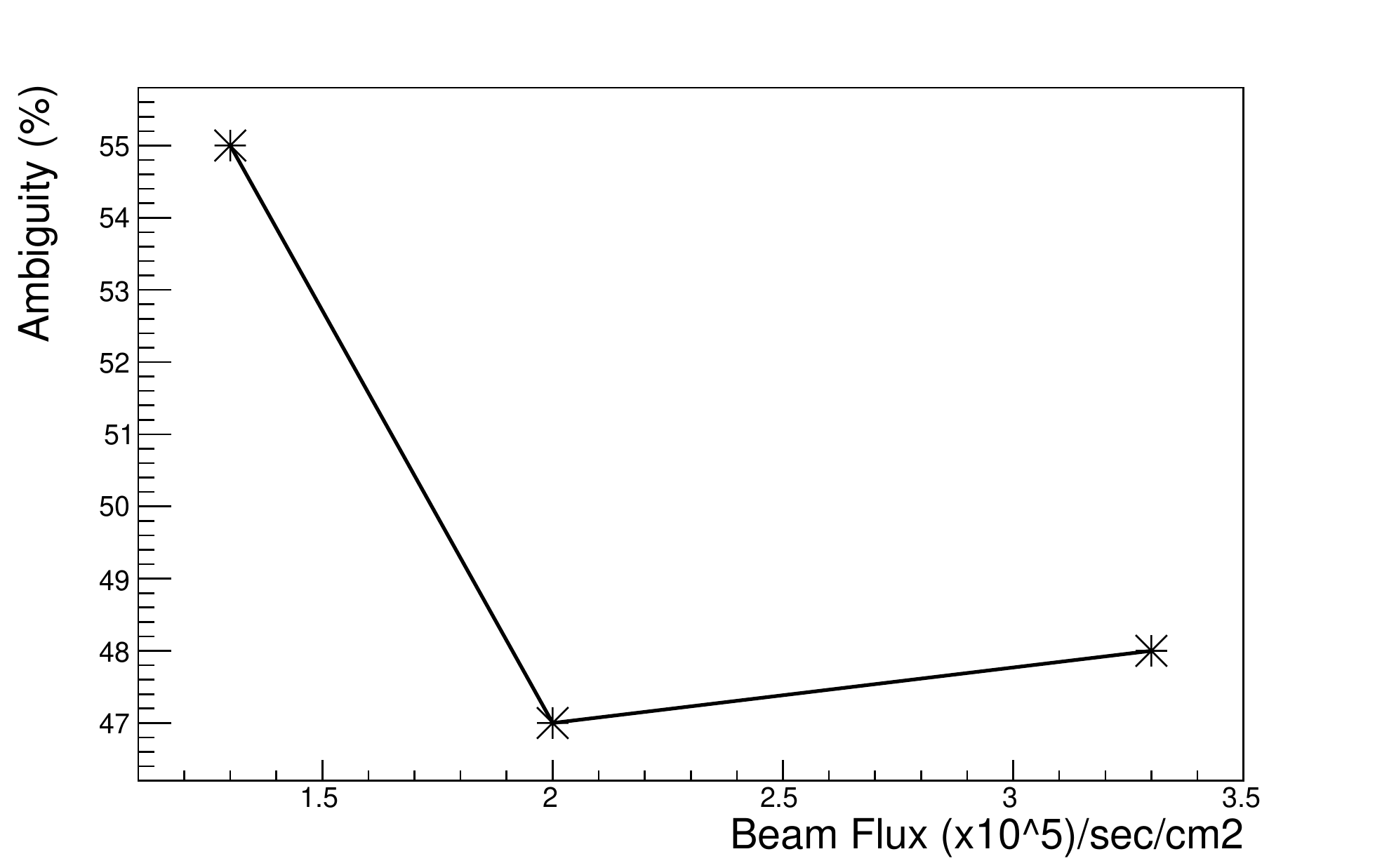}
\end{minipage}
\hfill
\begin{minipage}{.5\textwidth}
 \centering
 \includegraphics[scale=0.47]{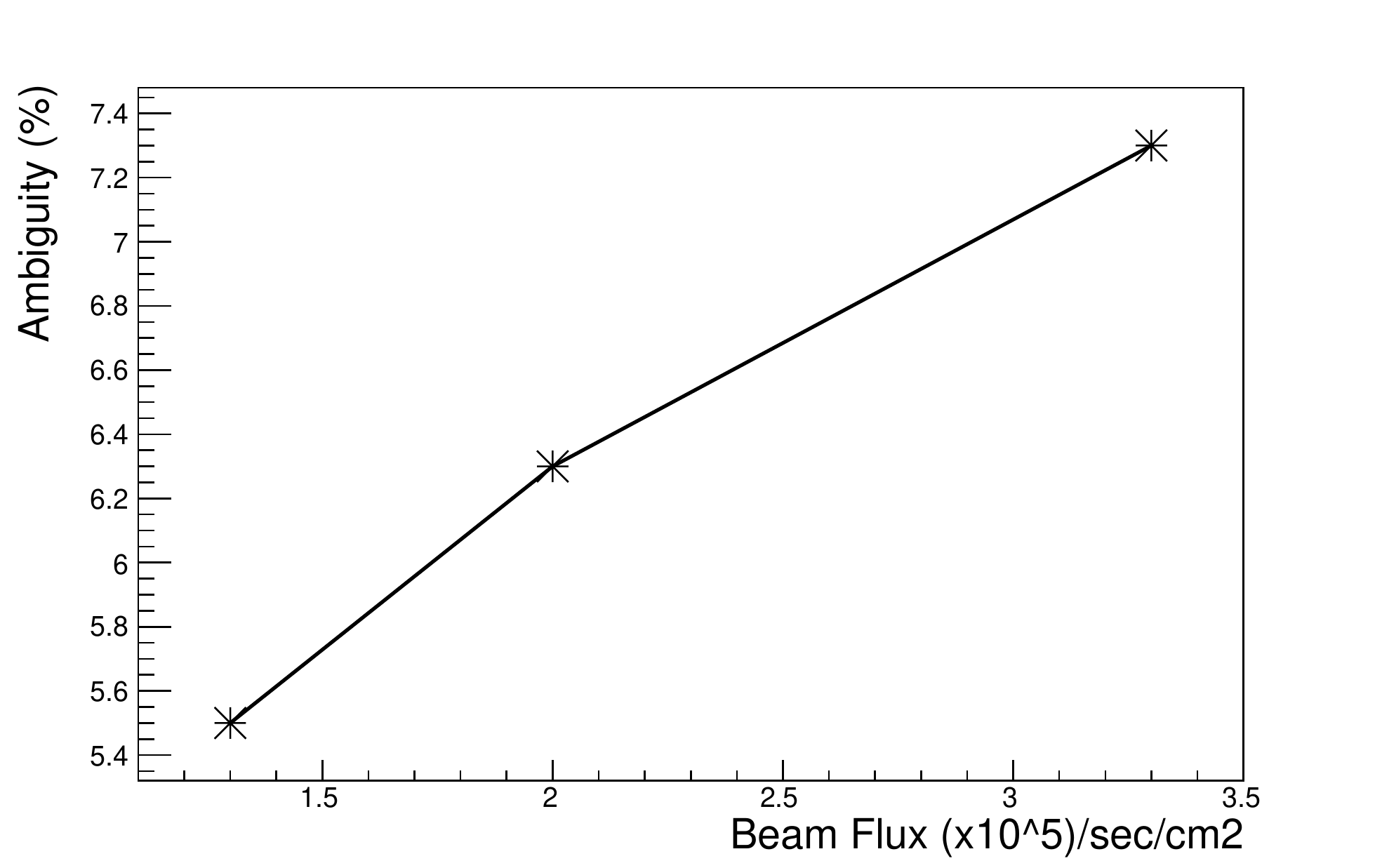}
 \end{minipage}
\caption{Probability of ambiguity ($\%$) due to cluster signal spread estimated for 1 particle hit events (left) and due to pileup events for 2 particle hit events (right).}
\label{clu_ambig}
\end{figure}

The multiplexing of the modules can cause some ambiguities in the signal reconstruction due to the loss of information. In fact if two particles enter the detector at the same time different combinations of channels may arise giving ``ghost" signal clusters. Smaller ``ghost" clusters with two or more strips may also arise if a signal cluster has a large spread ($>$ 2 strips). For example in our design, the $channel_{strip}$ combination for channels 0 and 7 are given by:
$0_{0}$  $0_{64}$  $0_{128}$  $0_{192}$  $0_{256}$ and 
$7_{7}$  $7_{65}$  $7_{163}$  $7_{253}$  $7_{287}$. 
Channels 0-7 are also connected to strips 0-7 apart from its other connections. So for a 7-strip wide signal cluster (not unlikely as will be shown in the following Section) between strips 0 and 7, the connection of channels 0 and 7 to the consecutive strips 64 and 65 will give rise to a fake combination of ``ghost" cluster. 
\paragraph{}
Here we show that one can substantially suppress the ``ghost" clusters by using the information from the integrated charge of the cluster and its size as proposed in \cite{procureur}. 
By listing all possible signal clusters on each plane that share the same readout channel, the cluster with the larger number of strips and with larger integrated charge is selected. In fact, all the others are results of fake combinations rather than real particle hits.
\paragraph{}
In order to estimate the level of ambiguity due to the spread of signal clusters larger than 2 strips, we compared 1 particle hit events before and after the cleaning. Fig. \ref{clu_ambig} (left) shows the fraction of events that were wrongly identified as having more than 1 cluster for 1 particle hit before the cleaning as a function of the beam flux. As expected there is no correlation between this probability and the beam flux.

\begin{figure}[H]
\centering
\includegraphics[scale=0.5]{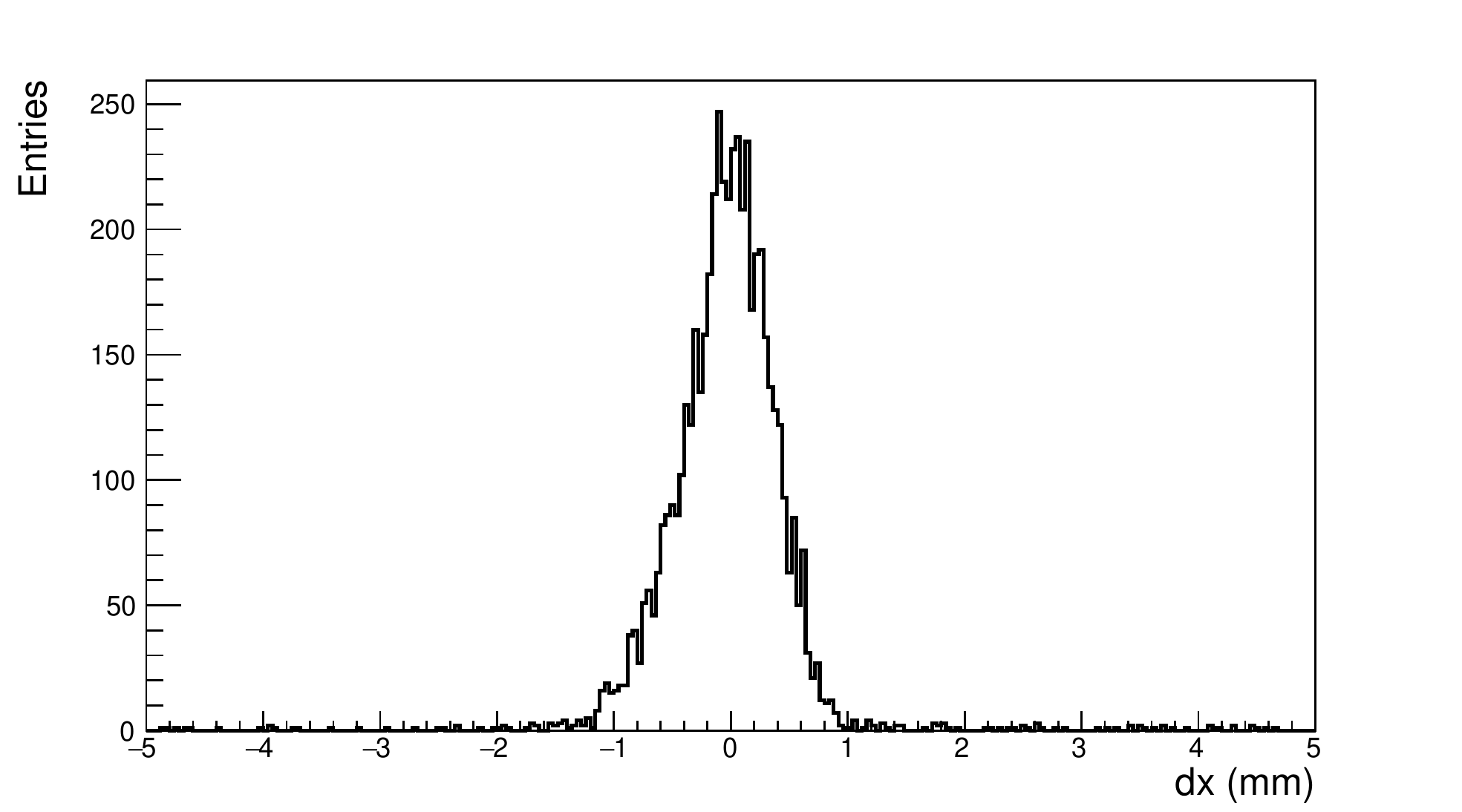}
\caption{Distribution of difference of cluster positions in each projection between MM 3 and 4 after selecting parallel tracks within the beam spot with MM 1 and 2 with energy 100 GeV $\pm$ 2 GeV selected with the ECAL.}
\label{mm34_del}
\end{figure}

To estimate the ambiguity due to pile up of particles, 2 particle hit events are compared before and after the cleaning. Almost 80 $\%$ of the 2 particle hit events give more than 2 signal clusters before the cleaning. Since, the fraction of 2 particle hit events range between 7 and 9 $\%$ of the total events depending on the flux, the probability of ambiguity due to the pile up is around 5-7 $\%$ for $\sim$ 100 kHz/(s$\cdot$cm$^2$) beam flux, as shown on the right side of Fig. \ref{clu_ambig}.  
\paragraph{}
To estimate the level at which a ``ghost" cluster was selected instead of the true signal cluster  with the method described above, we compared the position of the signal clusters on MM 3 and MM 4 after selecting a parallel incoming track within the beam spot using the position information from MM 1 and 2, having energy in the range 100 GeV $\pm$ $\sigma_{ECAL}$ where $\sigma_{ECAL}\sim 2$ GeV is the energy resolution of the ECAL. Fig. \ref{mm34_del} shows the distribution of the difference of cluster position on each projection between MM 3 and 4 for the same track candidate in an event. The distribution has a flat background with less than 2 $\%$ of the events with a difference larger than 4$\sigma_{MM}$ (where $\sigma_{MM}$ is the the MM hit resolution $\sim$ 100 $\mu$m, details given in Section \ref{resolution}). Systematic uncertainties due to misalignment between the MM modules are not taken into account. The estimation gives an upper limit to the level of wrong cluster identification after the cleaning method due to the multiplexing ambiguities. Thus we present an efficient way to limit the level of ambiguity decreasing substantially, from $\sim$ 50 $\%$ to $<$ 2 $\%$, the chance of wrong cluster identification using the cluster size, integrated charge and channel information.  With higher flux and pile up one can reduce the factor of multiplexing to limit the level of ambiguity further, depending on the acceptable level of ambiguities for the respective applications. One can also adapt the mapping of the channels to reduce the probability of ``ghost" clusters due to the charge spread for $k$-uplet clusters, where k is the average size of the clusters for the given application.
\subsection{Time Calculation}
\label{time}
\begin{figure}[H]
\centering
\includegraphics[scale=0.5]{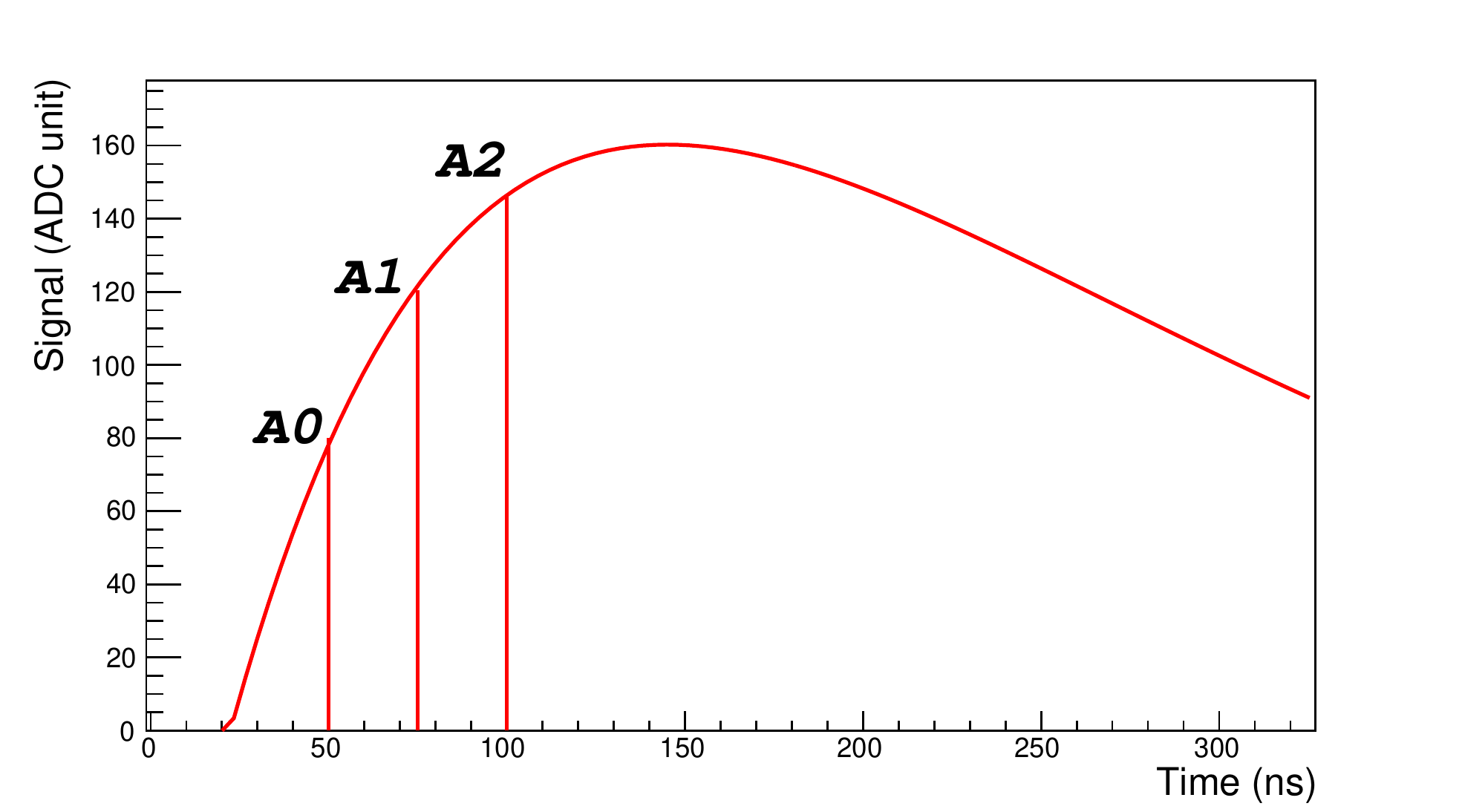}
\caption{Shape of signal sampled by a APV chip indicating the three samples 25 ns apart}
\label{sig_shape}
\end{figure}

\begin{figure}[H]
\begin{minipage}{.5\textwidth}
  \centering
\includegraphics[scale=0.48]{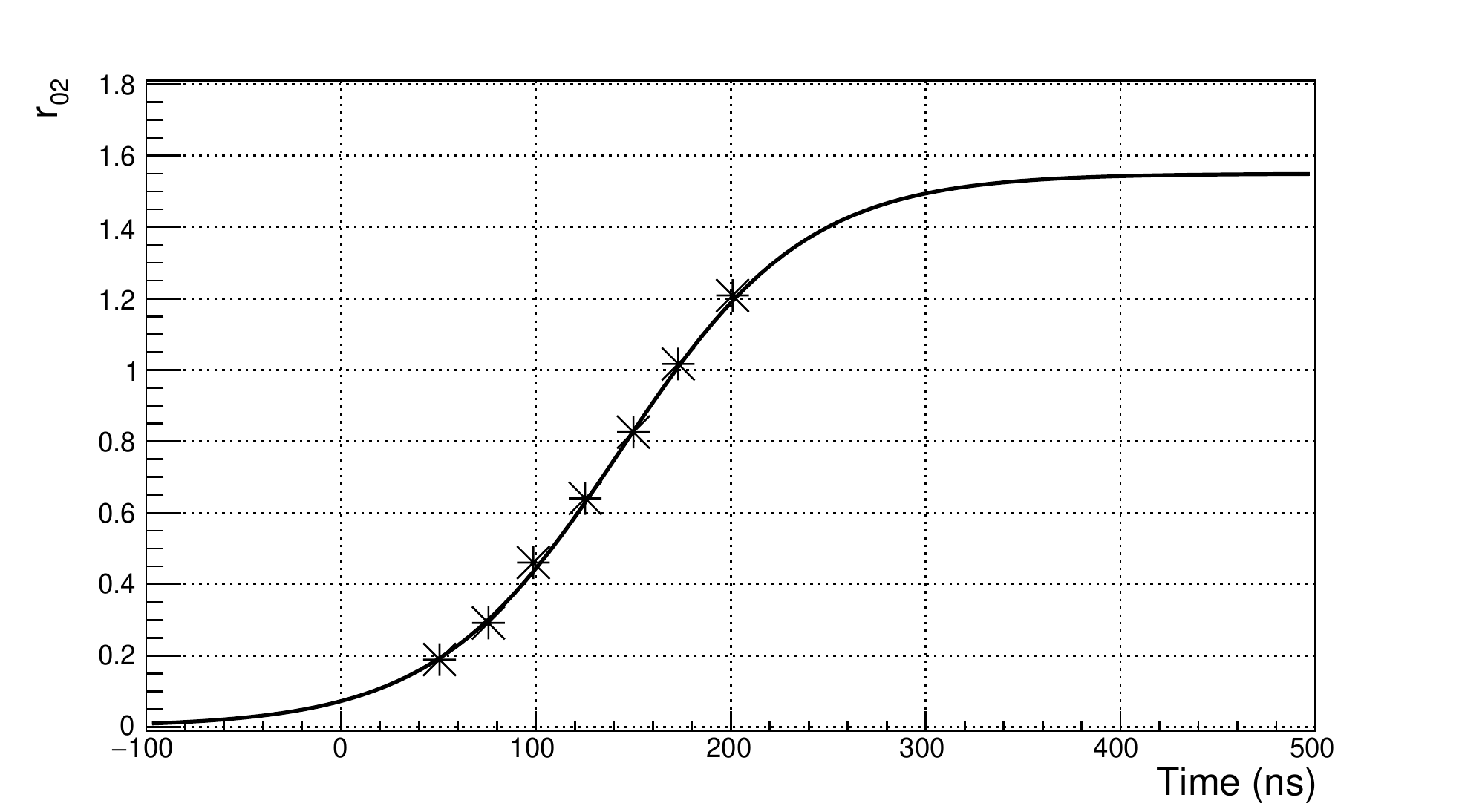}
\end{minipage}
\hfill
\begin{minipage}{.5\textwidth}
 \centering
 \includegraphics[scale=0.48]{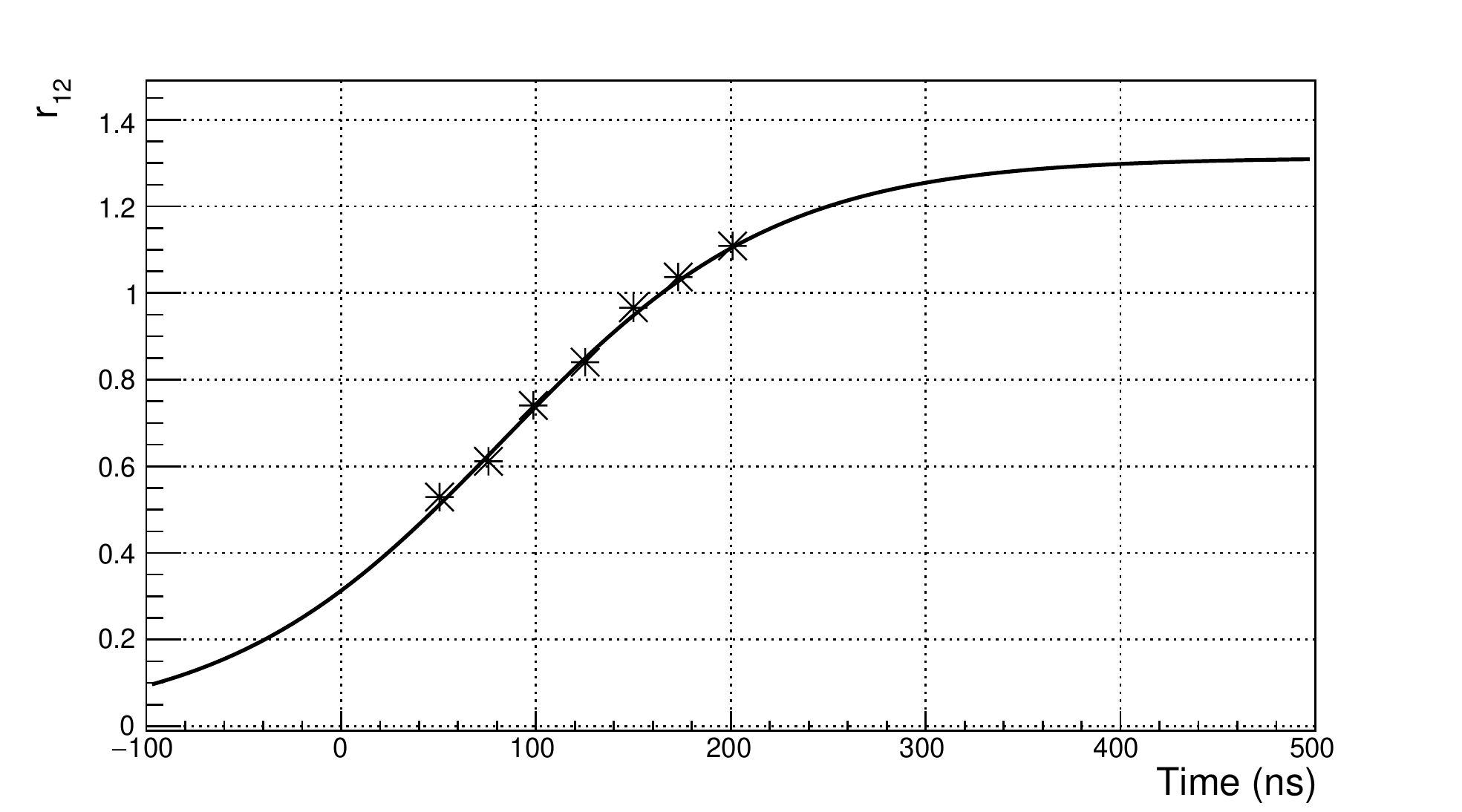}
 \end{minipage}
\caption{Ratio $r_{02}$ and $r_{12}$ as a function of the latency settings.}
\label{ratio_fit}
\end{figure}

 The timing calculation is based on the method used for the GEM detectors in the COMPASS experiment \cite{dipl_suhl}. The hit on a strip is defined when the weighted sum of the three analog samples from the APV25 chip (A0, A1, A2) are at least 3 standard deviations above the mean pedestal level. The latency between the trigger and the signal window is adjusted such that the three samples sit on the rising edge of the signal pulse as shown in Figure \ref{sig_shape}.   A scan between $\pm$ 75 ns (3 time sample units) with respect to the original latency setting was used to determine the rising edge. 
 To calculate the hit time for each strip and thus the time of the signal cluster, one defines the ratios
\begin{equation}
r_{02} = \frac{A0}{A2} \quad\text{and}\quad r_{12} = \frac{A1}{A2}
\end{equation}
that can be described by the function \cite{time_shape}
\begin{equation}\label{ratio}
r(t) = \frac{r_0}{(1+exp(\frac{t-t_0}{\tau}))}
\end{equation}
where $r$ is either $r_{02}$ or $r_{12}$.
 The three parameters $t_0$, $r_0$ and the slope $\tau$ along-with the covariance matrix, are found by fitting with Eq.\ref{ratio} the latency scan (Figure \ref{ratio_fit}).
The strip hit time can thus be calculated with
\begin{equation}\label{time_r}
t(r) = t_0 + \tau ln(\frac{r_0}{r}-1)
\end{equation}
The uncertainty on the ratios is given by: 
\begin{equation}
\sigma_{r_{02,12}} = (\sigma_n + \frac{1}{\sqrt{12}})\frac{\sqrt{A(0,1)^2 + A2^2}}{A2^2}
\end{equation}
$\sigma_n$ is the $\sigma$ of the pedestals of the channel reading the strip in the signal cluster. The additional $1/\sqrt{12}$ factor comes from the standard deviation of a standard uniform distribution on the strip connected to the channel.

 The calculations give two times per hit strip, $t_{02}^{i}$ and $t_{12}^{i}$, from the two ratios $r_{02}^{i}$ and $r_{12}^{i}$ with the corresponding parameters and errors for the i-th strip in the cluster. The errors on the individual ratio's time $\sigma_{t_{02,12}}^{i}$ (for the i-th strip) is calculated for all hit strips in the cluster and the cluster time is calculated using the weighted mean.

\section{Detector Performance}
The Micromegas gain was characterised with a radioactive source and then the modules were tested during the beam time of NA64 in October 2016 at CERN. During the four weeks beam run the performance of these multiplexed modules were checked for different beam intensities to establish their efficiency in high flux beam. 
The clustering algorithm for the multiplexed detectors, described above, was included in the data analysis.  
\subsection{Characterisation and Gain}
\begin{figure}[H]
\centering
\includegraphics[scale=0.35]{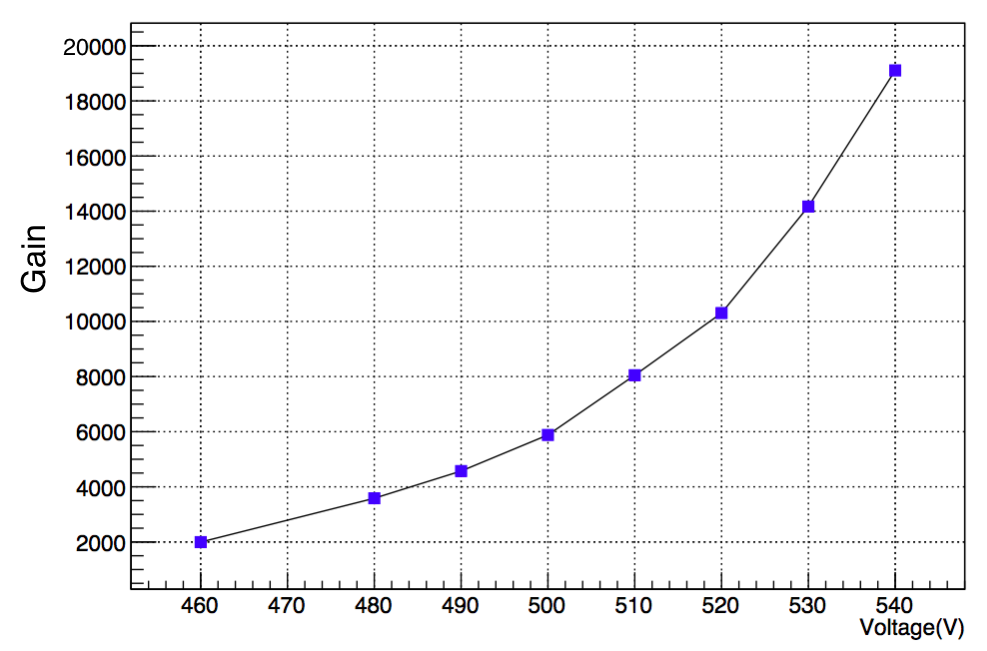}
\caption{Gain of a Micromegas module as a function of the amplification voltage}
\label{gain}
\end{figure}

The Micromegas detectors were first characterised with a radioactive $^{55}$Fe source to measure their gain. The gain, G,  is defined as the total number of electrons produced after amplification, per single incident electron in the gas volume as G = $\frac{N_{total}}{N_{prim}}$, where $N_{prim}$ is the number of electrons liberated from the ionisation of Argon in the drift region and $N_{total}$ is the total number of electrons after the amplification. The number of primary electrons is directly related to the nature of the gas and the energy of the incoming particle, E$_{x}$ =5.9 keV for $^{55}$Fe, as N$_{prim}$ = $\frac{E_{x}}{\omega_{i}}=223$ e$^{-}$, where $\omega_{i}$ = 26.5 eV/e$^{-}$ is the ionization potential of Ar-CO$_2$ (93-7 $\%$) \cite{gas_det}. So the gain of each detector was calculated by measuring the total current on the strips taking into account the rate of interaction from the source. The drift voltage was fixed at -300 V. The gain obtained as a function of the amplification voltage is shown in Fig. \ref{gain} for one module. The amplification voltage was kept below the spontaneous breakdown voltage limit of 570 V and the typical gain is about $2\times10^4$. Detailed reasoning of breakdown mechanisms in gas detectors is presented in \cite{breakdown}. 
\subsection{Hit detection efficiency}
\begin{figure}[H]
\centering
\includegraphics[scale=0.7]{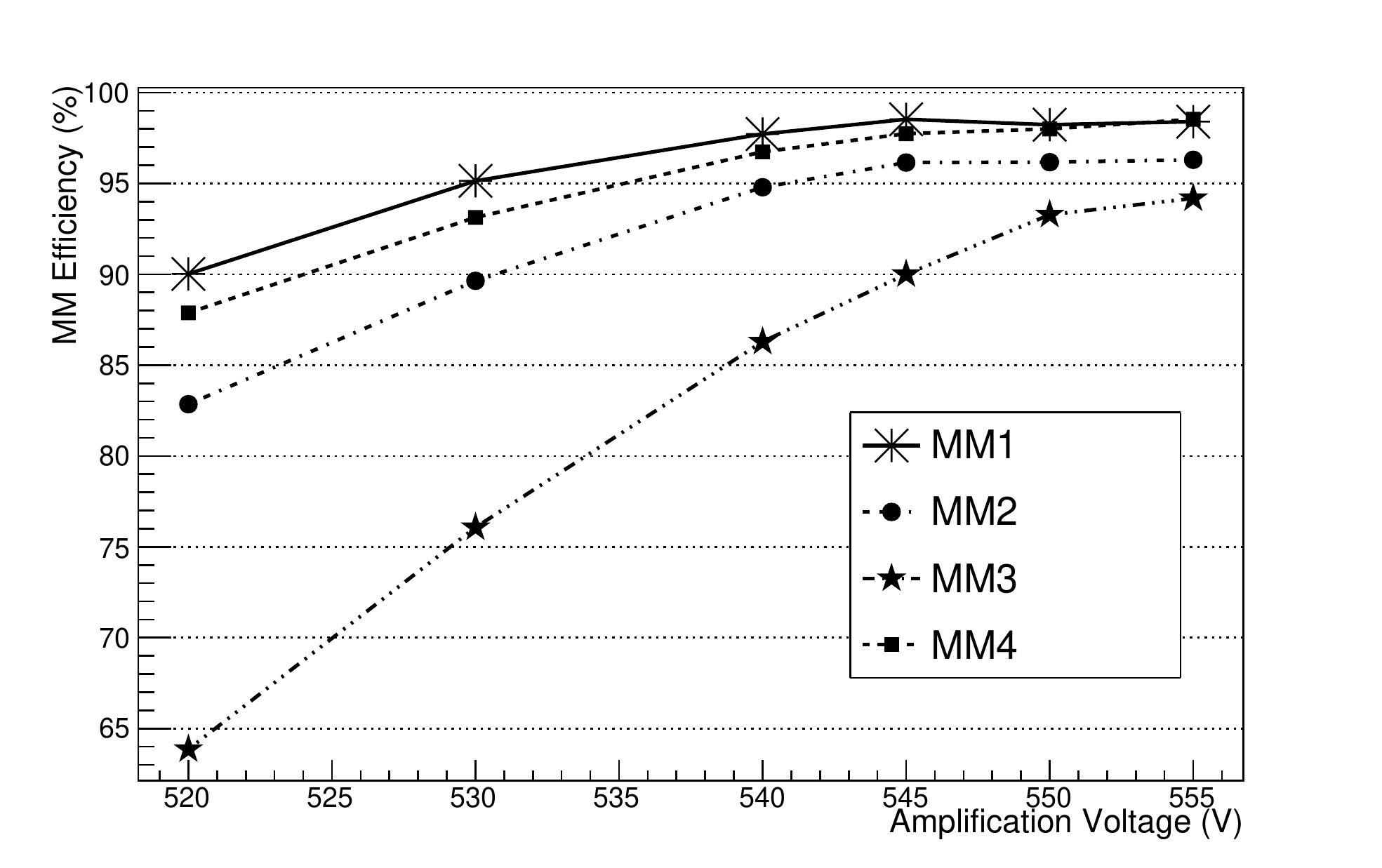}
\caption{Hit Detection Efficiency of the four MM modules as a function of the amplification voltage}
\label{MM_eff_volt}
\end{figure}

\begin{figure}[H]
\centering
\includegraphics[scale=0.7]{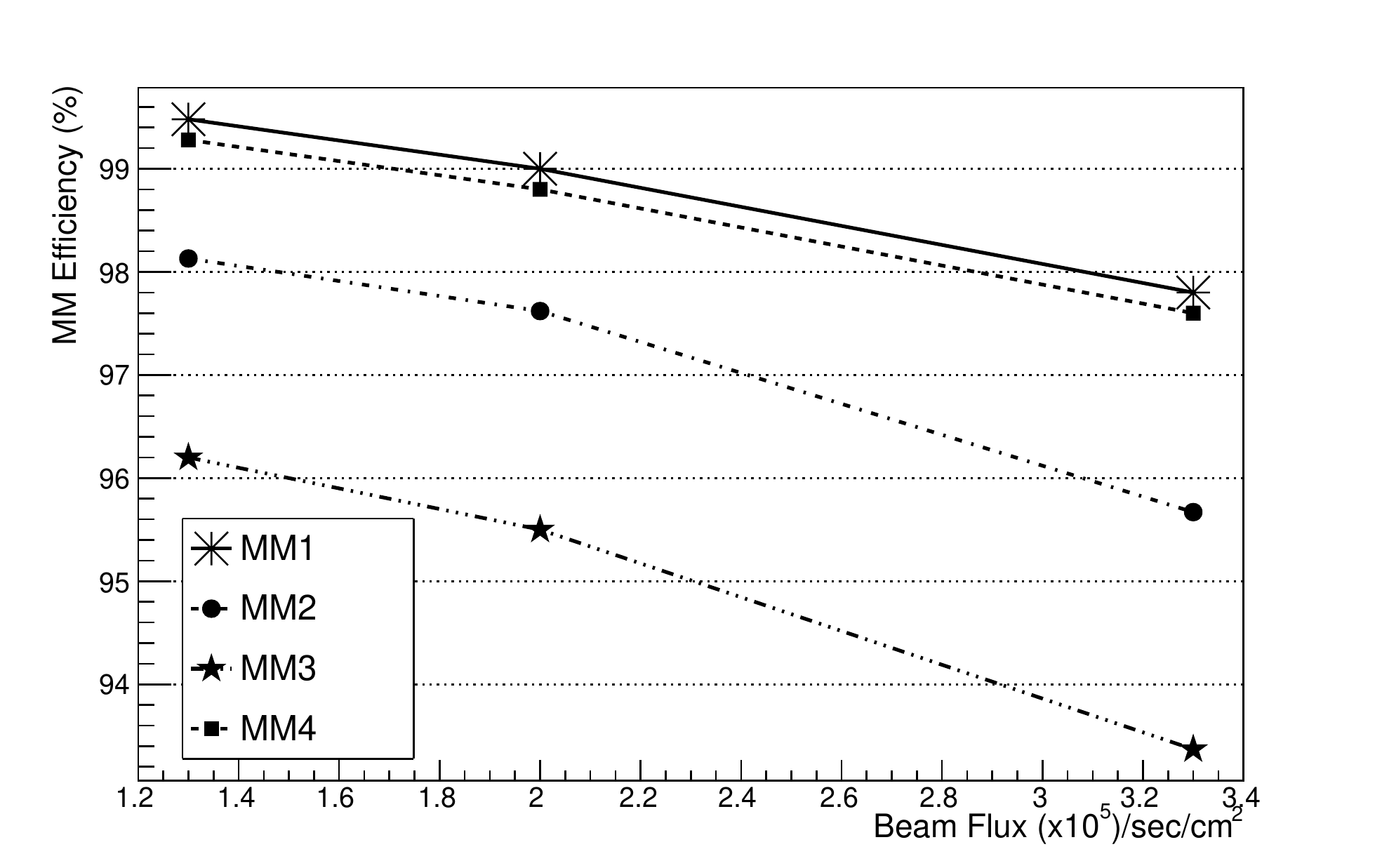}
\caption{Hit Detection Efficiency of the four MM modules as a function of the beam flux}
\label{MM_eff_rate}
\end{figure}

The hit detection efficiency, defined as the fraction of events with at least one signal cluster (on both X and Y plane) with respect to the triggered events, was measured as a function of the amplification voltage in the beam as shown in Fig. \ref{MM_eff_volt}. The Micromegas detectors were placed in the maximal beam intensity of 3.3 $ \times$ 10$^5$ e$^-$/(s$\cdot$cm$^2$). The efficiency for all 4 modules increases with increasing voltage, as expected with the increase of the gain, with the rate of increase falling as the voltage approaches the discharge limit. The Micromegas efficiency was also checked for different beam fluxes after fixing the amplification voltage at $\sim$ 555 V for MM 2, 3 and 4 and $\sim$ 545 V for MM 1, those being the voltage at which the respective modules were most hit efficient. Fig. \ref{MM_eff_rate} shows the efficiency of MM modules as a function of the beam flux. As one can see, for the maximal rate, the average hit detection efficiency of the four MM modules is $\sim$ 96 $\%$ with MM3 being the least efficient. It was found to be the most noisy detector from the pedestal distribution (larger pedestal standard deviation) of its electronic channels.
\subsection{Spatial Resolution}
\label{resolution}
\begin{figure}[H]
  \centering
  \includegraphics[scale=0.5]{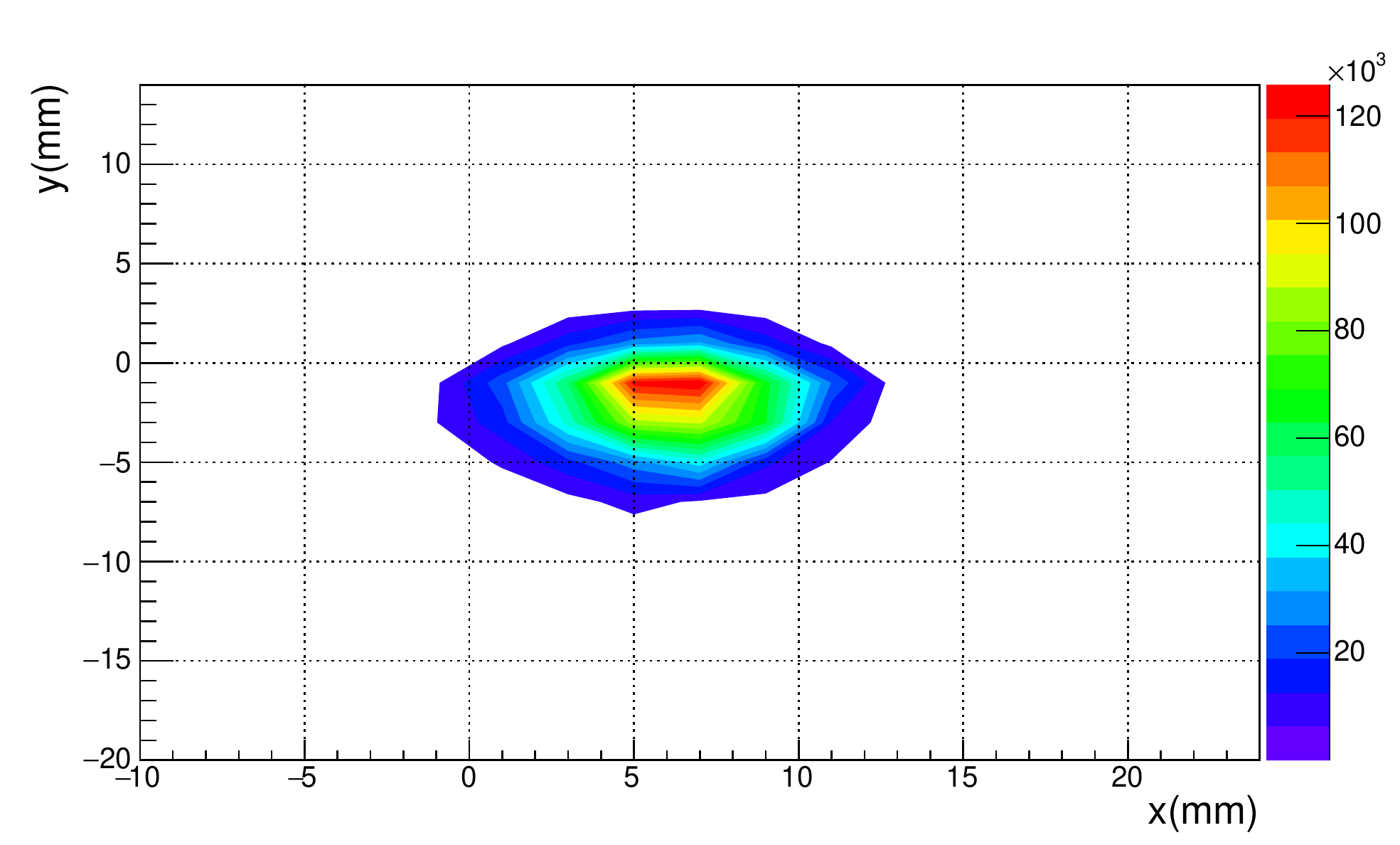}
\caption{Typical Beam spot on the four Micromegas modules. }
\label{beam_spot}
\end{figure}

\begin{figure}[H]
\begin{minipage}{.5\textwidth}
  \centering
  \includegraphics[scale=0.48]{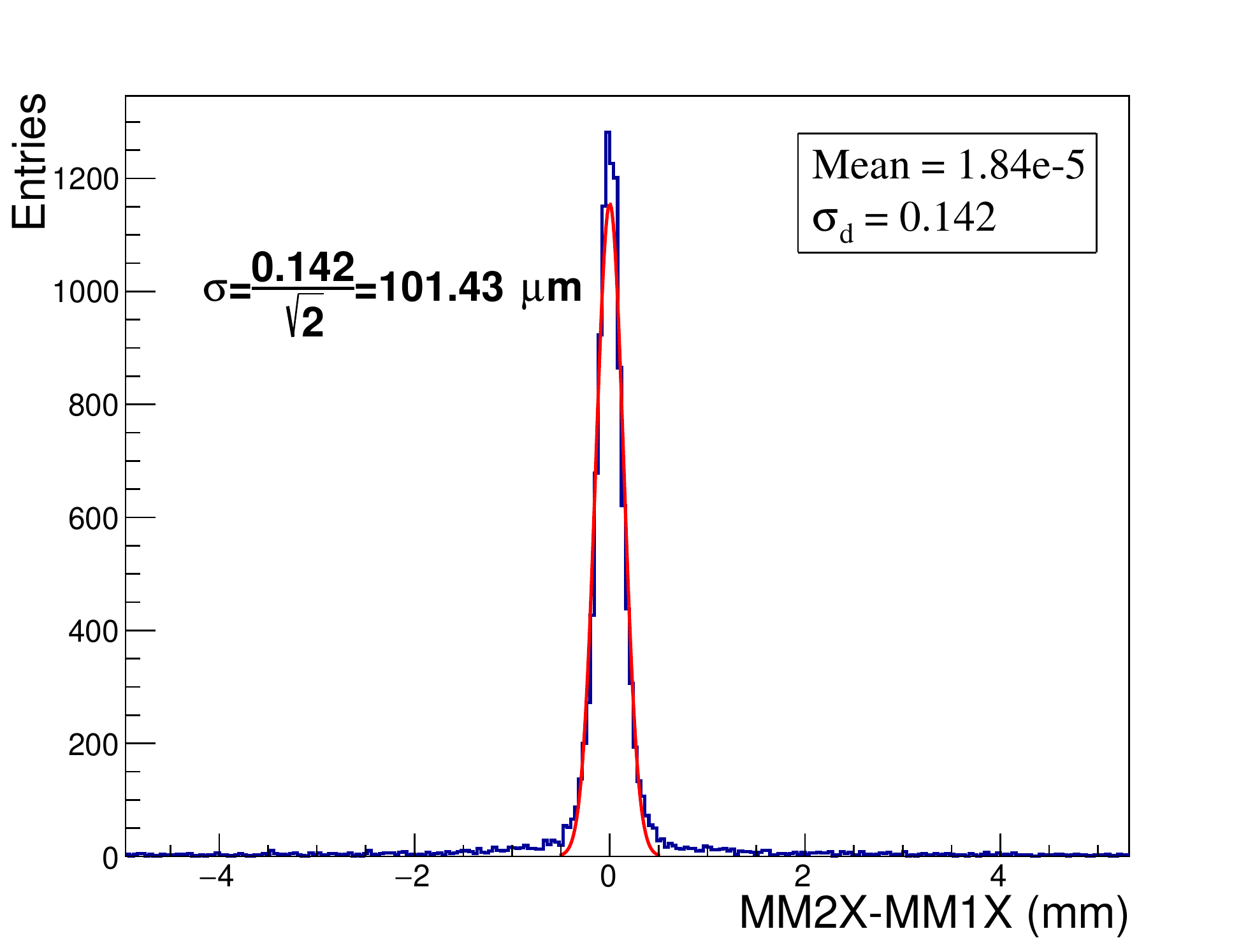}
  \label{m12x}
\end{minipage}
\hfill
\begin{minipage}{.5\textwidth}
  \centering
  \includegraphics[scale=0.48]{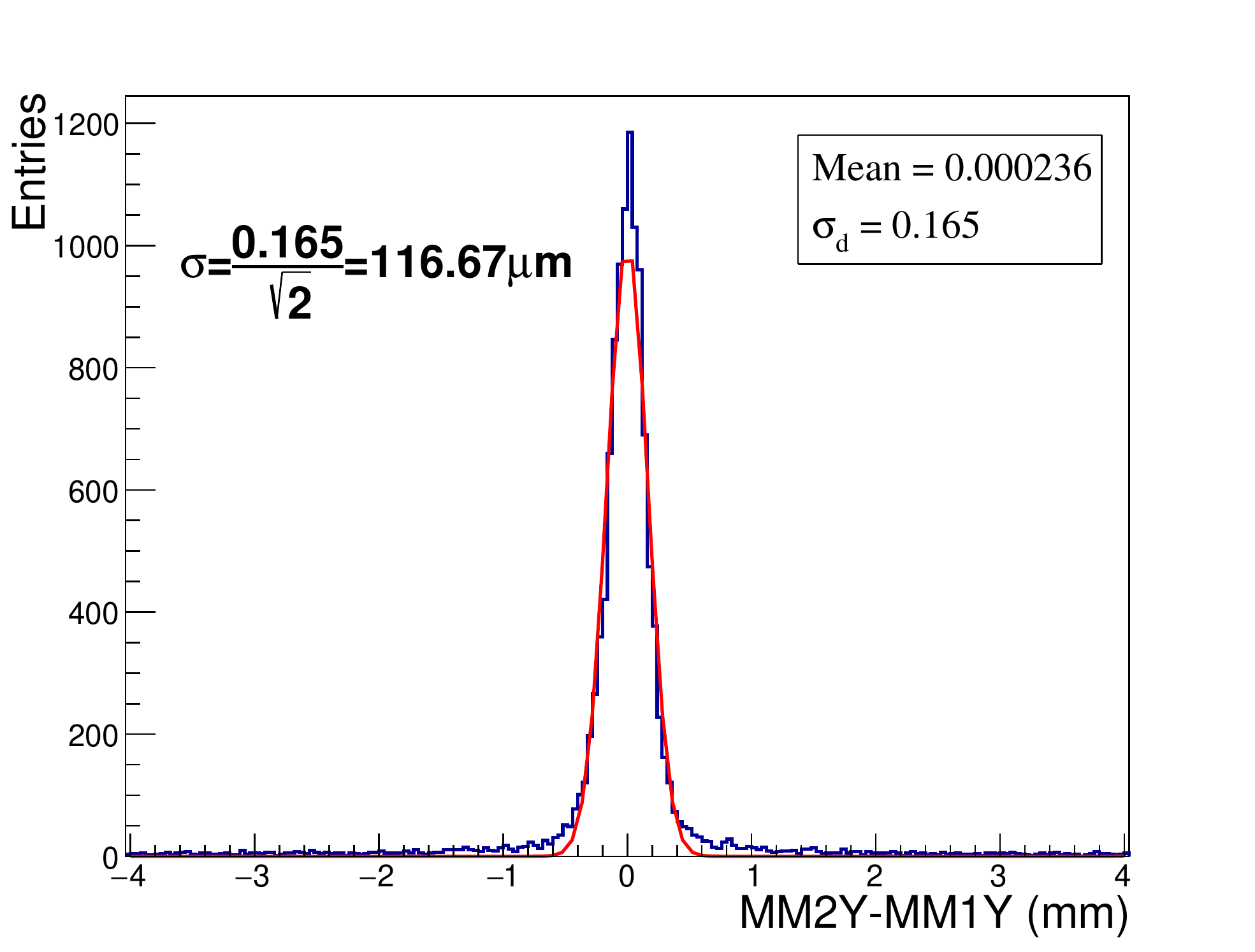}
  \label{m12y}
\end{minipage}
\caption{Distribution of difference of cluster positions in each projection between two MM modules. The black histogram is the data and the red line is a fitted Gaussian function with parameters ``$\sigma_{d}$" and ``Mean". Position resolution for each module $\sim$ 100 $\mu$m assuming same spatial resolution of each module}
\label{pos_res}
\end{figure}

\begin{figure}[H]
\begin{minipage}{.5\textwidth}
  \centering
  \includegraphics[scale=0.48]{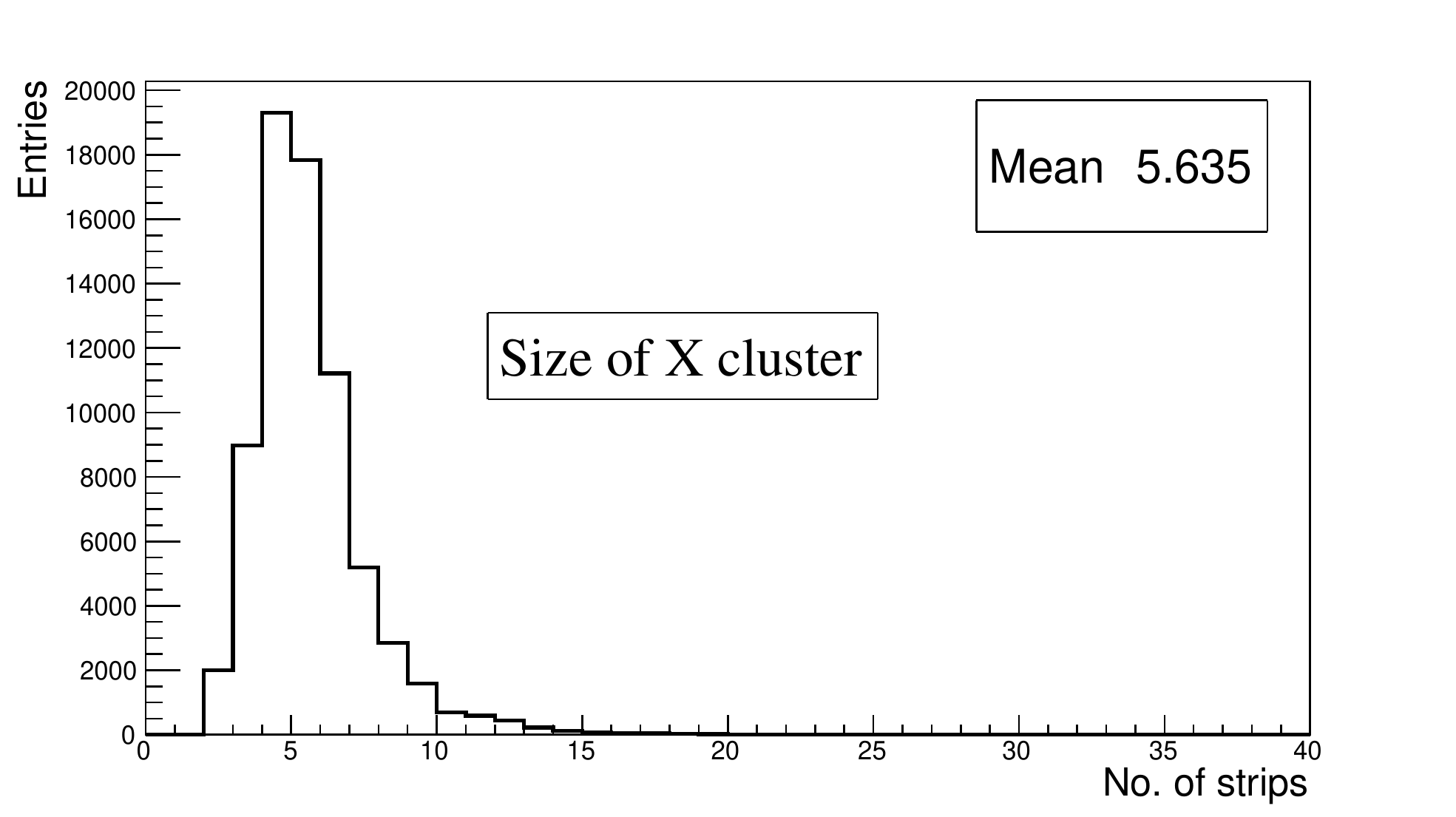}
  \label{x_size}
\end{minipage}
\hfill
\begin{minipage}{.5\textwidth}
  \centering
  \includegraphics[scale=0.48]{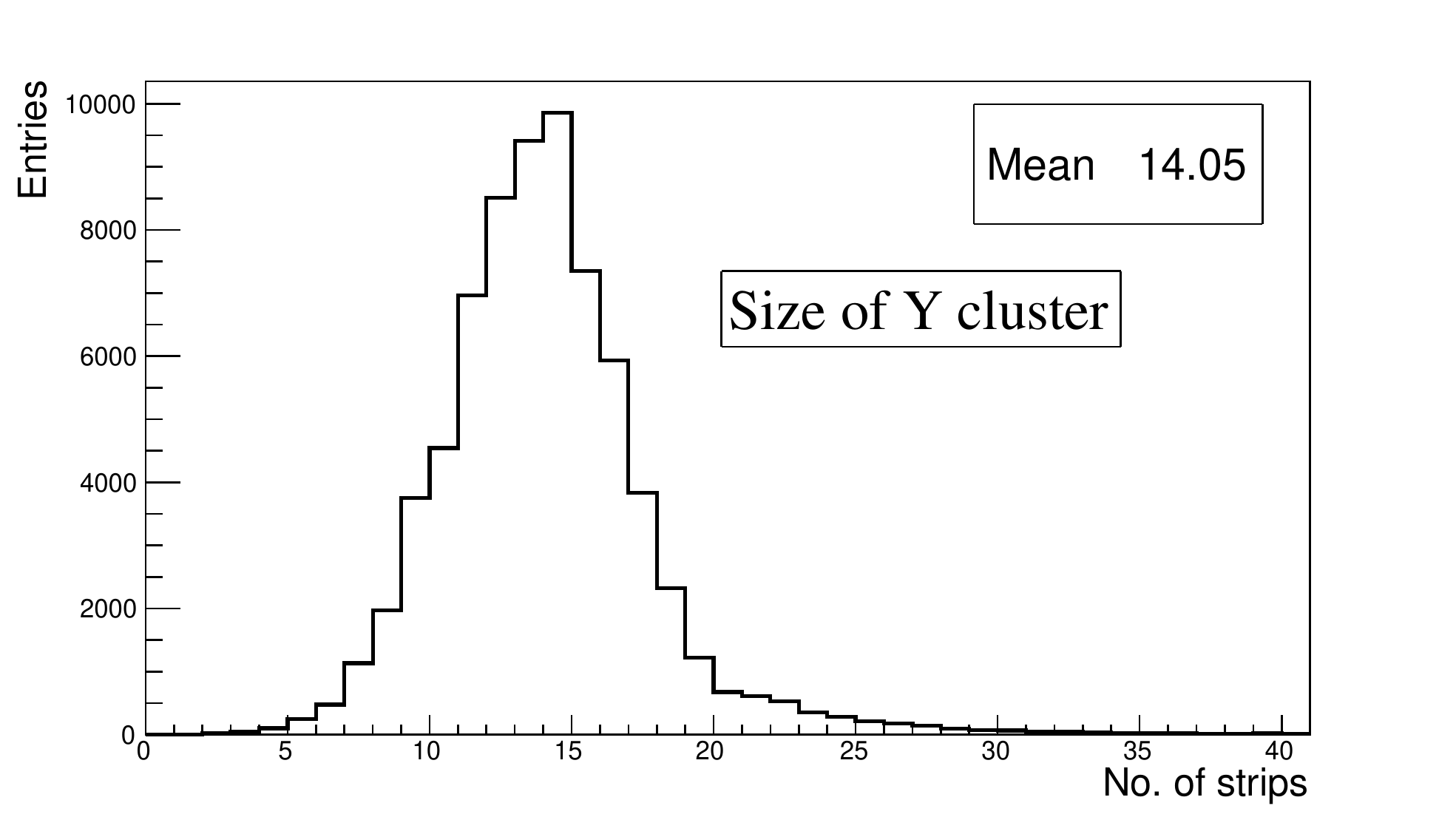}
  \label{y_size}
\end{minipage}
\caption{Size of clusters/plane in the Micromegas modules}
\label{cl_size}
\end{figure}

The spatial resolution of these modules were measured in the beam at the maximal beam intensity. Fig.\ref{beam_spot} shows the typical beam spot on the MM modules. The distribution of the difference between the signal cluster positions of the undeflected beam (without the magnetic field) on the MM modules is shown in Fig. \ref{pos_res}. The standard deviation of the distribution is a convolution of the spatial resolution of the two chambers. If we assume a parallel beam and the two chambers have same spatial resolution, $\sigma$, for a single chamber the resolution can be estimated to be $\sigma$ = $\sigma_{d}/\sqrt{2}$ $\sim$ 100 $\mu$m ($\sigma_{d}$ is the standard deviation of the distribution), in fair agreement with the theoretical value expected from the pitch of 250 $\mu$m of the strips which is $\sigma$ = $\frac{250}{\sqrt{12}}$ $\mu$m $\sim$ 72 $\mu$m. Uncertainties due to the beam divergence and relative misalignment between detector modules was not taken into account for this estimation. This shows that the modules not only are efficient but also have good hit resolution, not limited by the multiplexing ambiguities under such high intensities. Fig. \ref{time} shows the time distribution of each plane with respect to the time of scintillator S1. The resolution obtained is $\sigma_{t}$ $\sim$ 15 ns. The MAMMA collaboration tested Micromegas modules for the ATLAS New Small Wheel upgrade with APV chips sampling the entire signal shape and reported a resolution $<$ 10 ns in their test beam \cite{atlas_test}. Therefore, it should be possible to improve our timing, sampling the entire signal shape instead of just three samples as mentioned in Section \ref{sample}. The distribution of the size of clusters/plane is shown in Fig.\ref{cl_size} in units of number of strips. The difference in the cluster size in the X and Y plane is expected due to higher capacitive coupling to the Y plane than the X. 
\begin{figure}[H]
  \centering
  \includegraphics[scale=0.5]{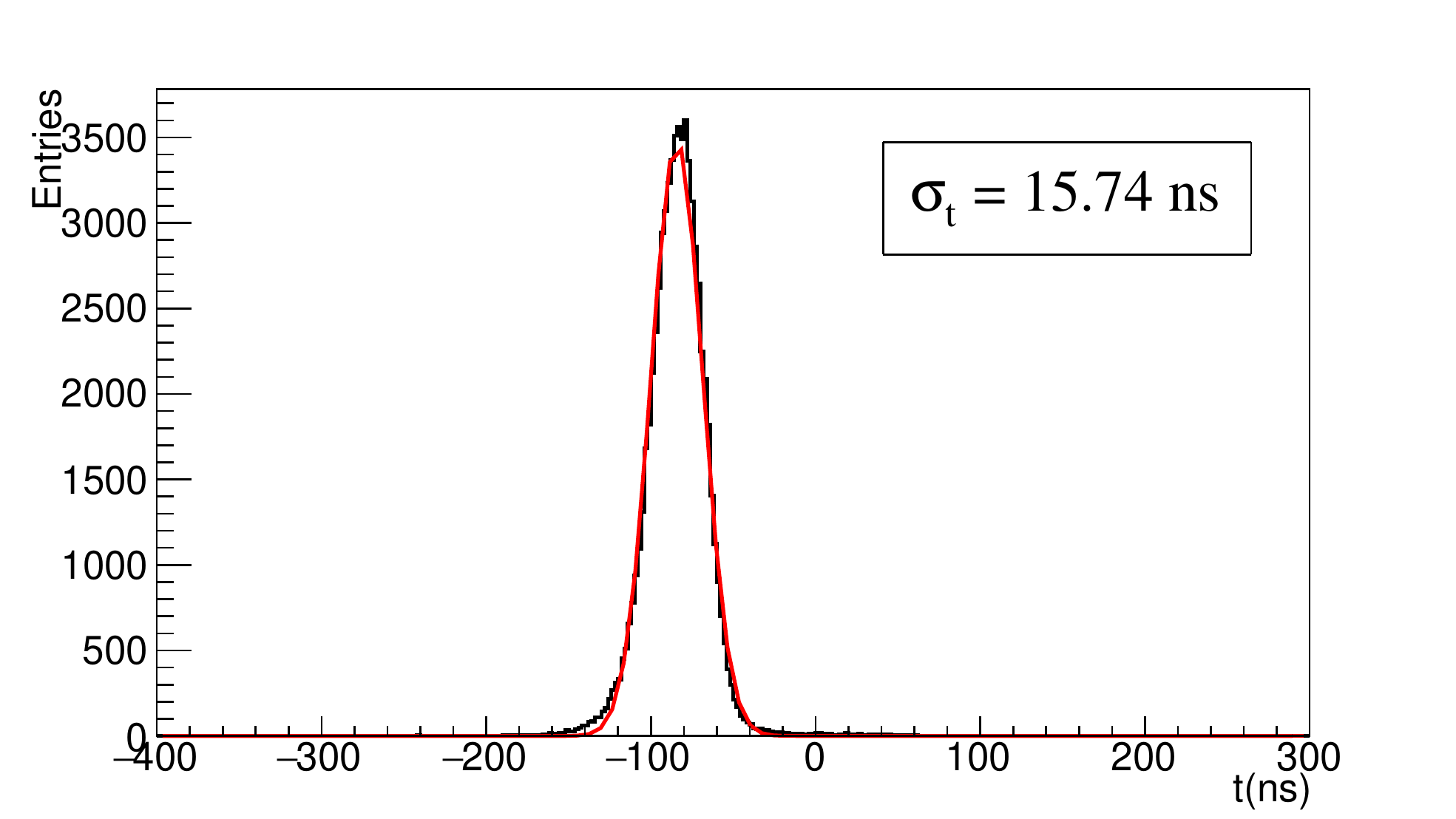}
\caption{Timing distribution of the Micromegas modules with respect to scintillator S1 time. The black histogram is data and the red line is a fitted Gaussian function with standard deviation $\sigma_{t}$}
\label{time}
\end{figure}
\subsection{Tracking in NA64 and suppression of low energy electron tail}
\begin{figure}[H]
\centering
\includegraphics[scale=0.7]{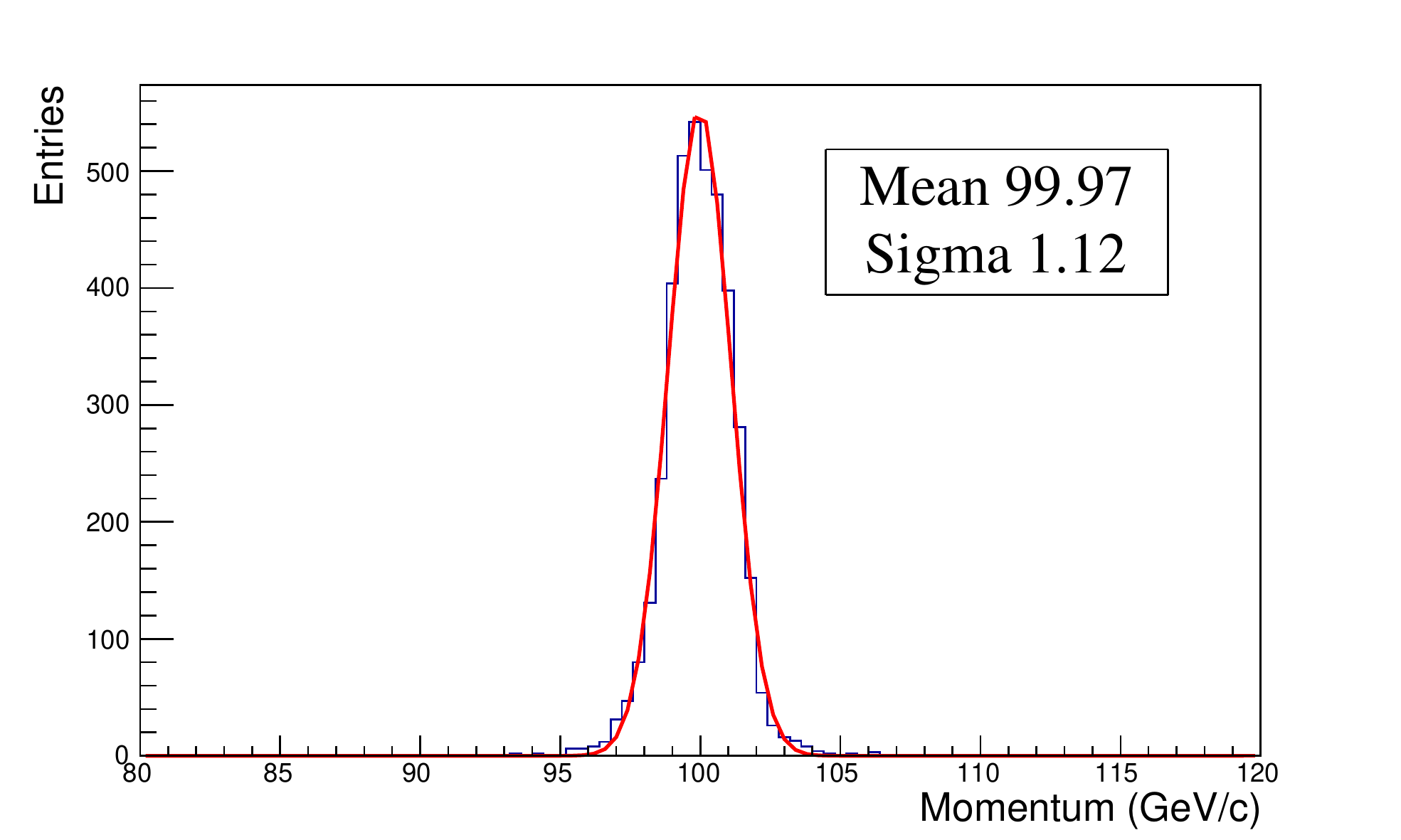}
\caption{Reconstructed momentum with the four Micromegas modules for a 100 GeV/c beam. The black histogram is data and the red line is a fitted Gaussian function with parameters ``Sigma" and ``Mean"}
\label{mom}
\end{figure}

The multiplexed modules were built for tracking in the NA64 experiment to suppress the low energy electron tail as mentioned above. Tracking of the incoming particles with the four modules was done under an integrated magnetic field of 7 T.m over two magnets with a combined length of 4.8 m. Fig.\ref{mom} shows the reconstructed momentum for a 100 GeV/c electron beam as obtained with the Genfit software (a generic track reconstruction framework for nuclear and particle physics \cite{genfit}). The resolution of the central peak is $\sim$ 1.1 $\%$ as shown in the plot with an efficiency of 85\%. To improve the tracking efficiency, the number of MMD stations that will be used in the next NA64 beam time will be doubled, i.e. 4 MMD's will be placed before and 4 after the bending magnets. This upgrade will result in an increased overall efficiency of 92\%.
 
\begin{figure}[H]
\begin{minipage}{.5\textwidth}
  \centering
\includegraphics[scale=0.47]{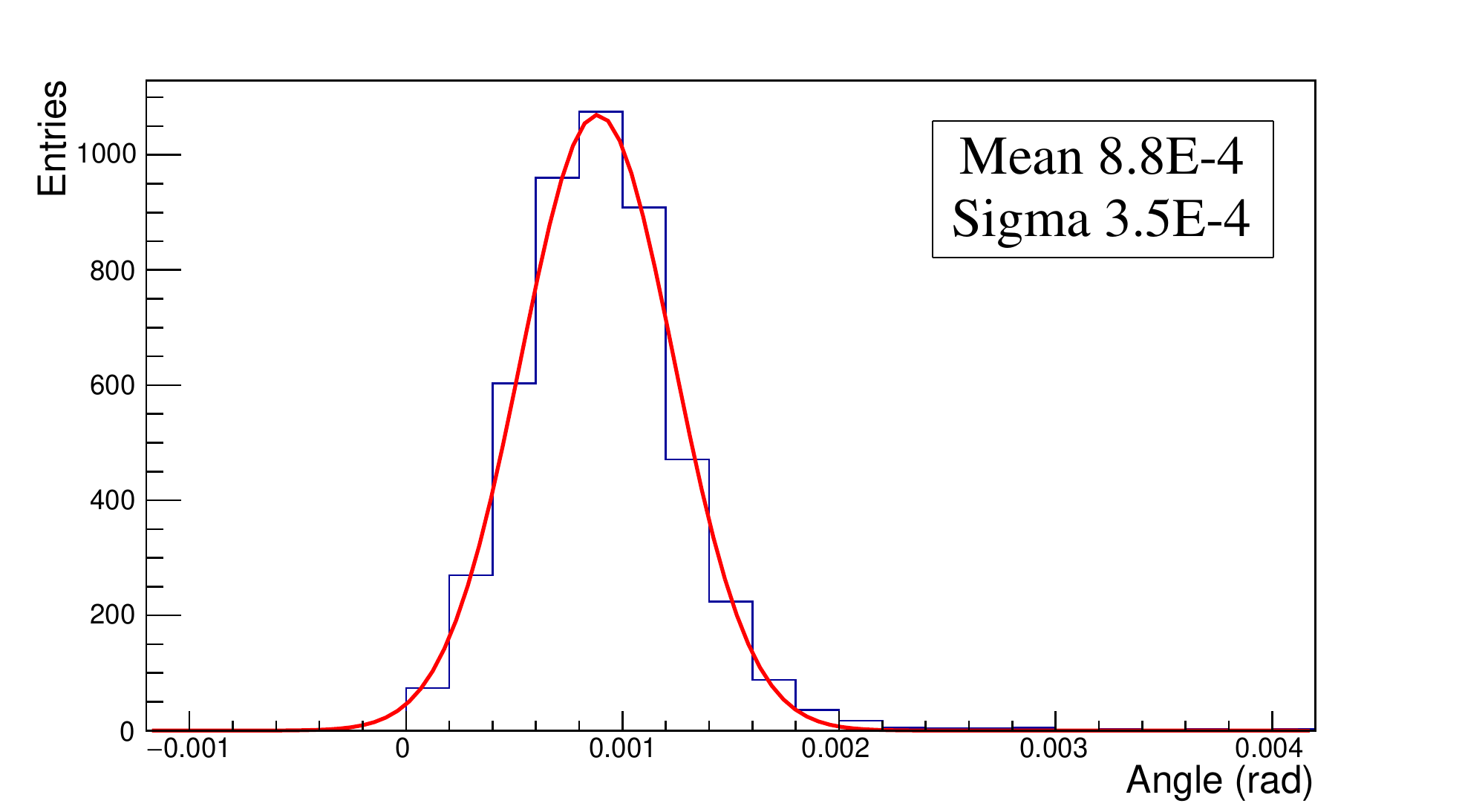}
\end{minipage}
\hfill
\begin{minipage}{.5\textwidth}
  \centering
\includegraphics[scale=0.47]{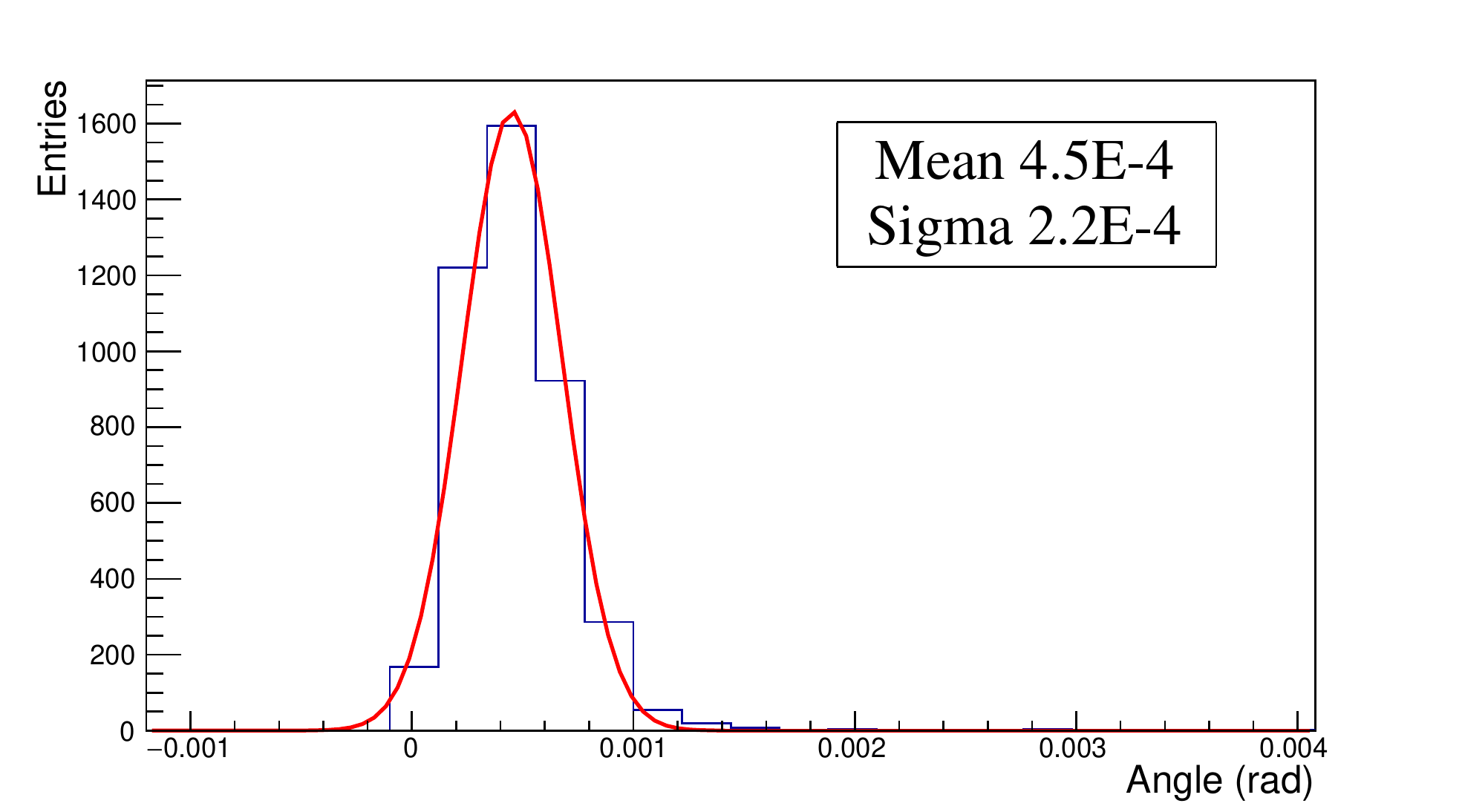}
\end{minipage}
\caption{Left: incoming particle azimuthal angle before the spectrometer measured with MM1 and MM2 with respect to the average beam axis. Right: outgoing  particle azimuthal angle measured after the magnets with MM3 and MM4 with respect to the average  deflected beam axis.   }
\label{ang_in_out}
\end{figure}

\begin{figure}[H]
\centering
\includegraphics[scale=0.5]{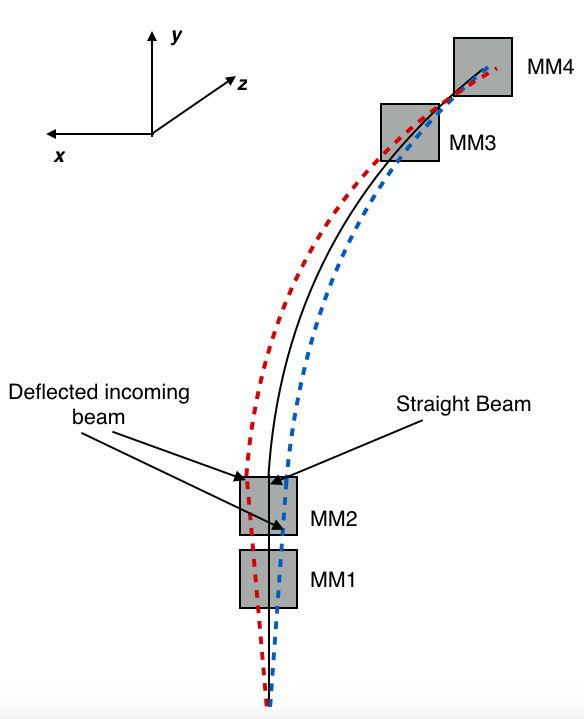}
\caption{Example of incoming beam deflection. Incoming angle is calculated with respect to the Z- axis}
\label{mom_ang_cart}
\end{figure}

\begin{figure}[H]
\begin{minipage}{.5\textwidth}
  \centering
 \includegraphics[scale=0.457]{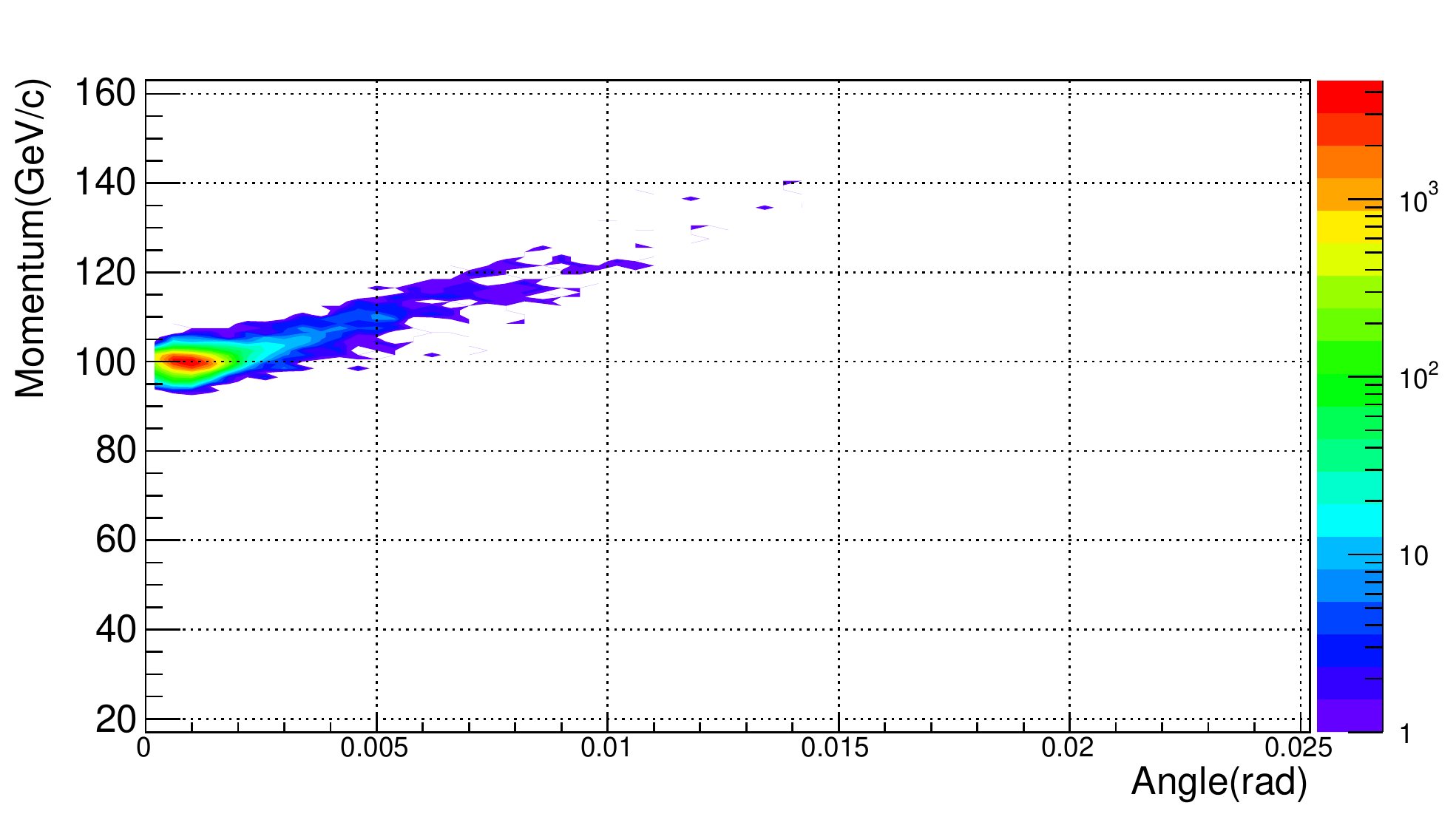}
\end{minipage}
\hfill
\begin{minipage}{.5\textwidth}
  \centering
  \includegraphics[scale=0.457]{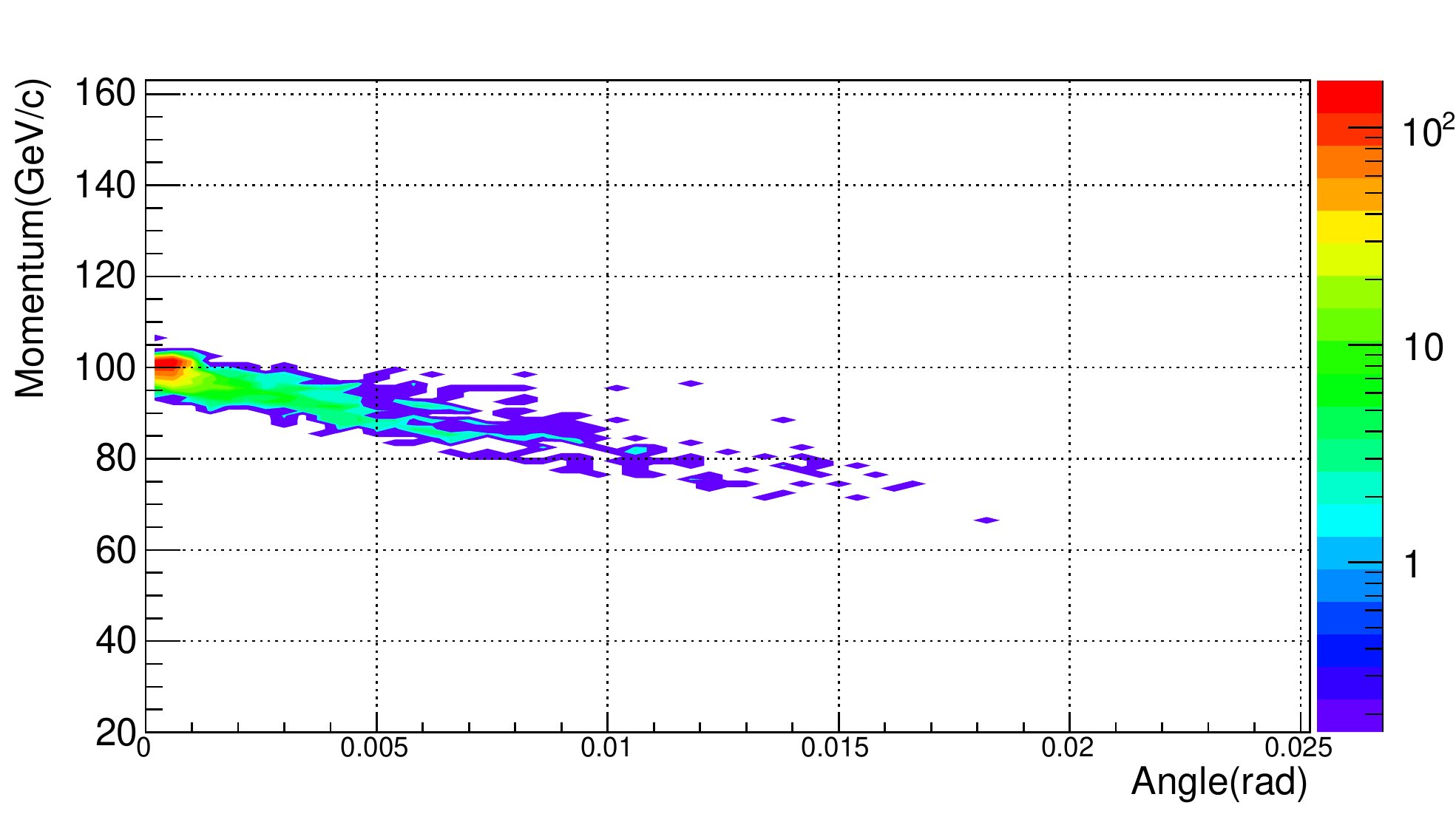}
  \end{minipage}
\caption{Momentum reconstructed as a function of the incoming angle for incoming particle deflected towards the negative x axis (left) and positive x axis (right).}
\label{mom_ang}
\end{figure}

The preparation of a collinear beam is a key point in NA64. The MMD were used to precisely measure the incoming angle of the  particles in order to tune and optimise the beam collinearity.  The angle was determined with an accuracy better than 1 mrad  allowing to reject large angle tracks and keeping the divergency of the beam within 1 mrad as shown in Fig. \ref{ang_in_out}. This accuracy allows to use the MMD also for the traversal scan of the ECAL hermeticity in order to look for non-uniformities.
In principle, higher and lower momentum particles should not be within the acceptance of the geometry unless they enter with large incident angles with respect to the primary beam direction. The reconstructed momentum was also checked as a function of the incoming beam angle. Fig.\ref{mom_ang_cart} shows a sketch of the incoming beam. The angle of the incoming beam with respect to the z-axis was calculated from the MM1 and 2 hit positions and the reconstructed momentum was plotted as a function of the angle as shown in Fig.\ref{mom_ang}.  As expected, when the initial deflection is in the negative x direction the reconstructed momentum is larger with increasing angle and when the initial deflection is in the positive x direction the reconstructed momentum tends to be smaller with increasing angle. Therefore, the incoming angle measured by the 2 MMs (MM1, MM2 in Fig.\ref{setup_cartoon}) upstream the magnet is a powerful tool to reject low energy electrons that are a dangerous background for the experiment \cite{PRL}. 
\section{Conclusion}
This paper presents the first results of Multiplexed Resistive XY Micromegas modules in a high flux beam performing with an average hit efficiency of 96 $\%$ per module for a multiplexing factor of 5. Multiplexed detectors are a novel idea to reduce the number of readout channels which prove to be very useful for the present need of particle physics experiments to cover large areas without compromising single hit resolution to perform precise tracking with reduced cost. So far the Multiplexed Resistive XY Micromegas modules have only been tested experimentally using Cosmic ray \cite{muon_tomo} for muon tomography studies with Micromegas based telescopes. Our work is the first to report on the performance of these detectors in a high flux beam environment. However, with multiplexing any grouping implies a certain loss of information. This is the reason why ambiguities can occur due to ``ghost" signals from fake combinations. 
Our measurements show that it is indeed possible to limit the ambiguities using information from cluster size, integrated charge and channel information with $<$ 2 $\%$ chance of wrong cluster identification, thus allowing for an efficient and reduced cost tracking detector at high rates. In this study we showed that the multiplexed technology can be efficiently extended to high rate environment for tracking experiments with a tracking resolution of 1.1 $\%$ for 100 GeV/c beam particles.
\section{Acknowledgements}
We gratefully acknowledge the help of S. Procureur and B. Vallage from CEA, Saclay, discussions with J. Wotschack from the ATLAS Group, and the support of E. Olivieri and F. Resnati from the RD51 collaboration. This work was supported by the HISKP, University of Bonn (Germany),  JINR  (Dubna),  MON  and  RAS  (Russia), the Russian Federation program ``Nauka'' (Contract No. 0.1764.GZB.2017), ETH Zurich and SNSF Grant No. 169133 (Switzerland),  and  grants  FONDECYT  1140471 and  1150792,  PIA/Ring  ACT1406  and  PIA/Basal  FB0821  CONICYT (Chile).  Part of the work on MC simulations was supported by the RSF grant 14-12-01430.   
We  thank the COMPASS  DAQ  group  and  the  Institute  for  Hadronic Structure  and  Fundamental  Symmetries  of  TU  Munich for the technical support.

\end{document}